\documentclass[sigconf, authorversion, nonacm]{acmart}

\AtBeginDocument{%
  }

\usepackage{float}
\usepackage{xspace}
\usepackage{graphicx}
\usepackage{enumitem}
\usepackage[normalem]{ulem}
\usepackage[dvipsnames]{xcolor}
\usepackage{makecell}
\usepackage{seqsplit}
\usepackage{tabularx}
\usepackage{array}    

\definecolor{tsuccess}{RGB}{22,95,130}
\definecolor{tfail}{RGB}{192,0,1}

\newcommand{\sayit}[1]{``\textit{#1}''}
\newcommand{\say}[1]{``#1''}

\newboolean{showcomments} 
\setboolean{showcomments}{false}
\ifthenelse{\boolean{showcomments}}
{\newcommand{\nb}[2]{
		\fbox{\bfseries\sffamily\scriptsize#1}
		{\sf\small$\blacktriangleright$\textit{\textcolor{blue}{#2}}$\blacktriangleleft$}
	}
	
}
{\newcommand{\nb}[2]{}}

\newcommand\ignore[1]{}

\newcommand{\tool}{\textsc{MagicCopy}\xspace}
\newcommand{\splitt}[1]{\texttt{\seqsplit{#1}}}

\begin{document}

\title{MagicCopy: Bring my data along with me beyond boundaries of apps}

\author{Priyan Vaithilingam}
\email{pvaithilingam@g.harvard.edu}
\affiliation{%
  \institution{Harvard University}
  \city{Boston}
  \country{USA}
}

\author{Elena L. Glassman}
\email{glassman@seas.harvard.edu}
\affiliation{%
  \institution{Harvard University}
  \city{Boston}
  \country{USA}}

\author{Nathalie Henry Riche}
\email{nathalie.henry@microsoft.com}
\affiliation{%
  \institution{Microsoft}
  \city{Redmond}
  \country{USA}
}

\author{Gonzalo Ramos}
\email{goramos@microsoft.com}
\affiliation{%
  \institution{Microsoft}
  \city{Redmond}
  \country{USA}
}

\author{Jeevana Priya Inala}
\email{jinala@microsoft.com}
\affiliation{%
  \institution{Microsoft}
  \city{Redmond}
  \country{USA}
}

\author{Chenglong Wang}
\email{chenwang@microsoft.com}
\affiliation{%
 \institution{Microsoft}
 \city{Redmond}
 \country{USA}}

\renewcommand{\shortauthors}{Vaithilingam et al.}

\begin{abstract}
People working with data often move their data across multiple applications, because they rely on these apps' complementing user experiences to best complete their tasks. Since traditional copy-and-paste approaches do not accommodate diverse table representations adopted by different apps, users spend considerable effort to reconstruct data formats and visual representations, making cross-app workflows costly. For example, when transferring a spreadsheet table with conditional formatting to a markup document, users spend substantial time translating its structure into appropriate tags and manually reformat color. 
This paper introduces MagicCopy, an AI-powered cross-app copy-and-paste, leveraging source and target contexts and user-specified instructions in natural language to automatically extract, parse, transform, and (re)format data from one app to another. 
In a study with sixteen participants, users quickly learned and applied MagicCopy to move data across three pairs of tools. Participants further explored diverse applications of MagicCopy to support more streamlined crossed-application interaction in their workflows.
\end{abstract}

\begin{teaserfigure}
  \centering
  \includegraphics[width=0.85\textwidth]{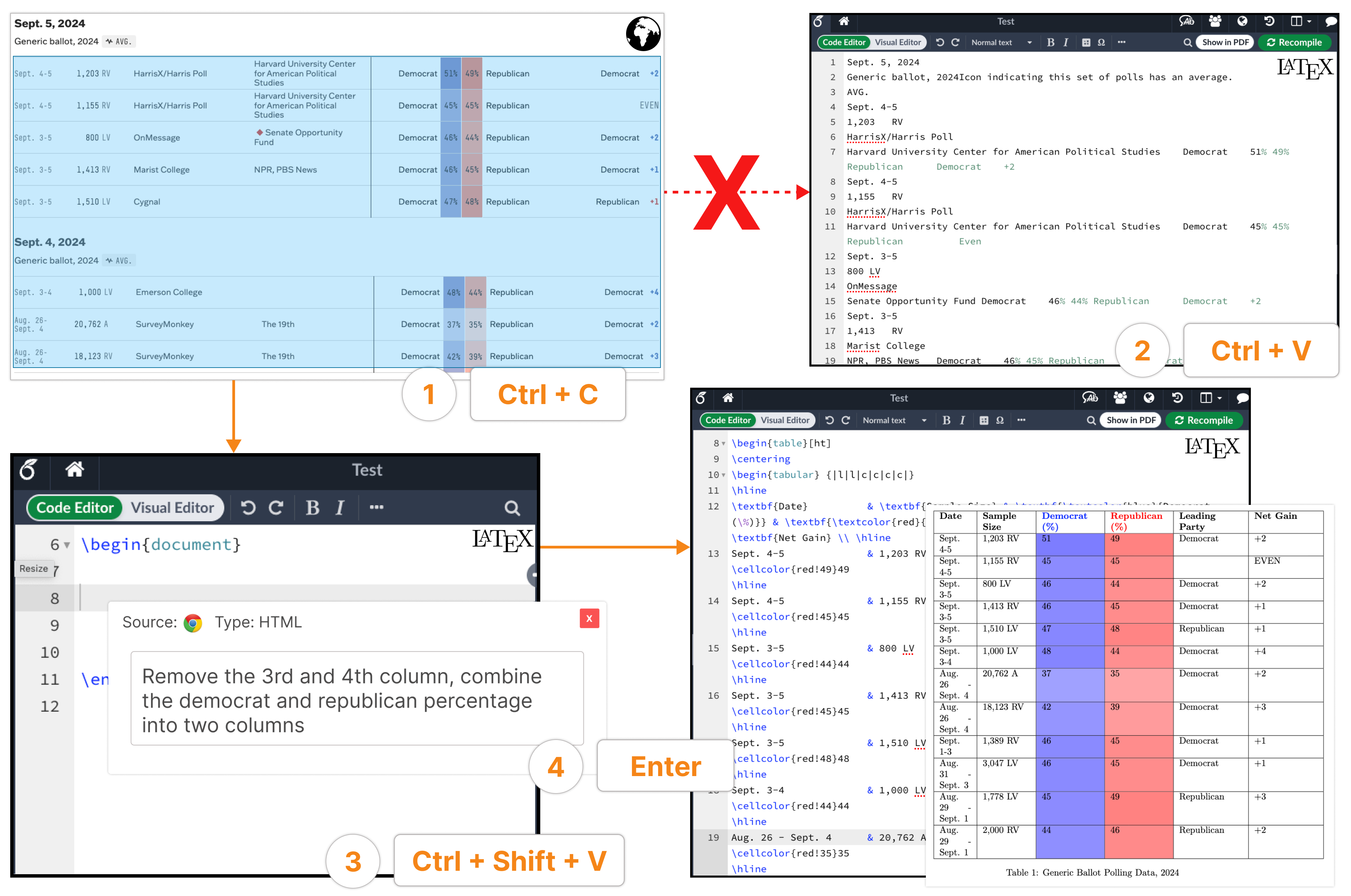}
  \caption{\tool in action. (1) The user copies an HTML table from the web to transfer it to LaTex with some modifications. (2) However, regular copy-paste fails due to incompatible data representation. (3) The user triggers a paste using our tool \tool, and provides an instruction to perform an additional table transformation while pasting. (4) \tool successfully pasted the correctly transformed table while preserving the format from the original source.}
  \Description{Shows an image of \tool's example use-case. (1) The user copies an HTML table from the Web to transfer it to LaTex with some modifications. (2) However, regular copy-paste fails due to incompatible data representation. (3) The user triggers a paste using our tool \tool, and provides an instruction to perform an additional table transformation while pasting. (4) \tool successfully pasted the correctly transformed table while preserving the format from the original source.}
  \label{fig:teaser}
\end{teaserfigure}


\maketitle

\section{Introduction}

People working with data, such as data scientists, analysts, and researchers, often need to move their data across multiple applications, because these applications provide complementing experiences that allow the user to more effectively complete tasks in different stages of their workflow~\cite{Woodruff2019DataTA, Reif2006SemanticC, stolee2009revealing, zhang2020data}. For example, to analyze questionnaire data from a user study, a researcher often starts with a spreadsheet tool like Excel, as its table user interface (UI) allows them to easily navigate the data and fix small user input errors. Then, to perform statistical analysis, the researcher might move the data to a programming environment like Jupyter Notebook, where the programming UI lets them quickly explore different analysis approaches and create visualizations. Finally, to publish the results, they need to move and format the final data in a LaTeX editor, where typesetting language allows them to flexibly adjust its presentation. 

Despite the benefits of working with multiple applications, users face the challenge of transferring data across applications, where they often need to \emph{transform and (re)format} the data before it is usable within the destination application. This challenge arises as the result of the distinct ways data is represented across the applications (e.g., HTML tables in websites, typeset tables in Latex, data arrays in Python Notebook, custom table objects in Excel, etc.) and simple copy-and-paste actions or moving data by exporting/importing common file formats (e.g., \textsf{.csv} files) often do not carry the structure and formats of the table from the source application to its destination. For example, as illustrated in Figure~\ref{fig:teaser}, as the user copies a web table and pastes it into Excel, the result loses its original structure, let alone its format (e.g., conditional formatting) and the user needs to manually typeset the table to restore its structure. These manual efforts are minimally annoying for users, and they sometimes require substantial efforts that become a barrier for users to use certain applications together despite their potential benefits (e.g., moving a financial table represented as a pivot table with multiple levels of header in Excel to R for statistical analysis, which requires tidy data formats).

Recent advancements in AI create an opportunity to address this challenge, especially the ability of Large Language Models (LLMs) to transform data between different formats based on the user's natural language instructions. Today, users have been using generative AI (GenAI) tools like ChatGPT as an ``intermediate stop'' between applications, where they copy the source data into ChatGPT, instruct it to transform data, and then paste the results into the destination tools; developers have also developed new AI-assisted data transformation tools within different applications (e.g., Office Copilot~\cite{office-copilot}) to assist data transformation. However, these approaches do not yet provide a ``streamlined'' experience for users to move data. With intermediate AI tools, users not only need to spend efforts learning different UIs and context-switching away from the main applications to get help from AI, but also have to prepare verbose prompts to describe their data transformation and formatting goals, as the data copied from the original application also loses its metadata and format information when the user moves the data from the source application to the AI tool.

To address these challenges, our key insight is that, instead of building new standalone AI tools for data transformation, we can \textbf{attach AI assistance to a universal action, copy-and-paste, with a transient user interface, supporting instrumental interaction paradigm}~\cite{beaudouin2000instrumental}. In this paradigm, the copy-and-paste action acts as an instrument, enabling users to seamlessly invoke AI assistance in situ. This integration minimizes workflow interruptions and cognitive load, as users interact with a well-integrated instrument that supports data transformation and reformatting across diverse applications. In this paper, we introduce \tool, an AI-powered universal copy-and-paste action, for users to transform and move their data across applications. As illustrated in Figure~\ref{fig:teaser}, as the user copies data (with shortcut \texttt{ctrl-c}) and pastes it into a new application (\texttt{ctrl-shift-v}), the user can provide a brief explanation of the paste requirements in a simple dialog box, and \tool automatically generates code to transform and format the data based on both the user instructions and the source and destination application contexts, to make it ready to use at the destination. \tool's action-centric design has the following two benefits. First, since copy-and-paste is a universal action that generalizes across applications, users can invoke and apply \tool in a wide range of applications in a uniform manner, even if these applications have very different User Interfaces (UIs) and data representations. Second, unlike application-centric AI tools (e.g., ChatGPT, Office Copilot) that can only access inputs provided by the user, \tool can keep track of context across applications (triggered when copy and paste actions are taken), thus the user does not need to write a verbose prompt to re-describe the format of the original data or destination format requirement---they only need to write a brief prompt to elaborate additional transformation goals they want to apply.

We build \tool as an OS-level action triggered along with the user's copy (\texttt{ctrl-c}) and paste (\texttt{ctrl-shift-v}) inputs. To make \tool context-aware, we introduce the \textbf{implicit context tracking}, where \tool stores a superset of information (e.g., HTML tags, attributes from Excel tables) besides the raw data from the source application around the data when ``copy'' is triggered, and this context is refined together with the destination contexts only when ``paste'' is triggered. This crucial design ensures that \tool does not lose information, especially formatting and structural information, upon copy, despite the fact that \tool cannot anticipate where the data will be moved to and what metadata will be needed by the user and the destination application. With the merged contexts, \tool then asks the large multi-modal model (LMM) GPT4o~\cite{gpt4o} to generate Python code to transform the data. Here, instead of directly asking the LMM to transform the clipboard content (represented as strings), the system transforms the data with generated code to reduce the risk of LMM hallucinations and to support larger input data that may not fit into the LMM's context window. As we will demonstrate, this design allows \tool to parse and clean semi-structured data (e.g., copying output data in a Jupyter notebook cell into Excel), transcribe image data to structured text (e.g., a web table image to markdown), paste data as a code snippet (e.g., into Python or R notebook), transfer a formatted table from PowerPoint to Latex while preserving formatting, etc., and support other diverse sets of applications and data formats. 

To understand how users could work with \tool to move data across applications and the potential applications of \tool in users' daily workflows, we conducted a user study with 16 participants. In this study, we first asked participants to complete two challenging tasks (that cannot be achieved with simple copy-and-paste) where they needed to move data between two tools assigned by us, and then we let them freely apply \tool to any tool and dataset they were interested in, in an open exploration format. All participants successfully applied \tool to complete the main tasks with an average of only 1.1 attempts per task. We documented the full list of scenarios explored by the participants in the open exploration: participants successfully used \tool to complete 47 out of 51 scenarios, despite many scenarios being outside of our expectations when we designed \tool (e.g., cleaning and formatting gene data with a regular expression along with pasting). Participants stated that \tool's \emph{``quick and dirty''} interaction approach has the potential to improve their workflow across applications.



\section{Usage Scenario}
This section describes scenarios in which people transfer and transform data across multiple applications, highlighting the challenges and manual effort involved in these tasks. We then discuss how \tool streamlines and simplifies the data transfer process.

\subsection{Motivating Scenarios}
\subsubsection*{\textbf{Scenario 1: Moving data across apps for different stages of user workflow}}

Alice is an HCI researcher working with user study data that she wants to analyze and write an academic paper on the findings. This workflow of Alice requires multiple steps---data cleaning, data analysis and aggregation, and report generation. Since these steps require different processes, Alice wanted to use the tools that she liked best for each of the steps. She opens the data in Excel (see Figure~\ref{fig:scenario1}~(1)) since it allows her to easily spot and correct spelling and typing errors directly in the cells. Once the data is clean, she wants to move to Jupyter Notebook to use Pandas, a familiar library, to manipulate and aggregate the study data. After computing a summary table in Jupyter Notebook, she plans to move this table into her Overleaf document for her paper. 

However, Alice's seemingly simple three-step workflow becomes complicated because the tools do not seamlessly interact. Moving data from Excel to Jupyter Notebook requires downloading the data as a CSV file, writing Python code to load the CSV file as a Pandas dataframe, and formatting the Pandas dataframe (such as fixing the index and converting the data to the right type). Then, copying the summary table from Jupyter to Overleaf introduces another challenge, as she needs to convert the HTML table into Latex format, which usually requires her to manually enter it herself. However, recently she has been using ChatGPT. She copies the data to ChatGPT window and gives specific instructions on converting it to a LaTex table --- adding an extra step to her workflow.

\subsubsection*{\textbf{Scenario 2: Switching to another tool for quick editing of data}}

Ethan is writing a blog post on the impact of COVID-19 across various regions. He has a simple table in markdown format embedded in his draft, showing the region, population, and the number of recent cases for each area. Ethan realizes that to better emphasize certain trends in his blog, he needs to add a new column—\say{Cases per 1,000 people}— derived from the \say{No. of cases} and \say{Population} columns and sort the table by this new column.

Since his blog editor does not have any tools to manipulate data, Ethan copied his data to Excel, where he used a simple formula ``\text{= (C2 / B2) * 1000}'' to compute the 4th column and used Excel's GUI to sort the data in the descending order of the 4th column. Once the computations and sorting are complete, Ethan tries to copy the updated table from Excel into his blog editor. However, he realizes that the table is pasted in plain text without markdown formatting, forcing him to manually typeset the table row-by-row. The time saved by using Excel for computations is now lost in manually converting the table back into markdown.

\subsubsection*{\textbf{Scenario 3: Syncing data + metadata across  apps}}
Amy, an analyst, is creating a PowerPoint presentation that includes a table with the results of her model comparisons across various benchmarks. Additionally, she also has to compose a summary article for her company's website featuring the same results. She discovers that transferring her highly customized PowerPoint table to HTML format requires manual entry of each cell's unique styling -- consuming the next hour of her time. The situation worsens when her manager requests a color scheme adjustment in the PowerPoint table to align with the company's color palette --  another hour lost. To prevent further changes to the presentation, Amy decides to avoid her manager for the rest of the day.

\subsubsection*{\textbf{Scenario 4: Moving  table with a different format for easier analysis}}
\begin{figure}
    \centering
    \includegraphics[width=0.85\linewidth]{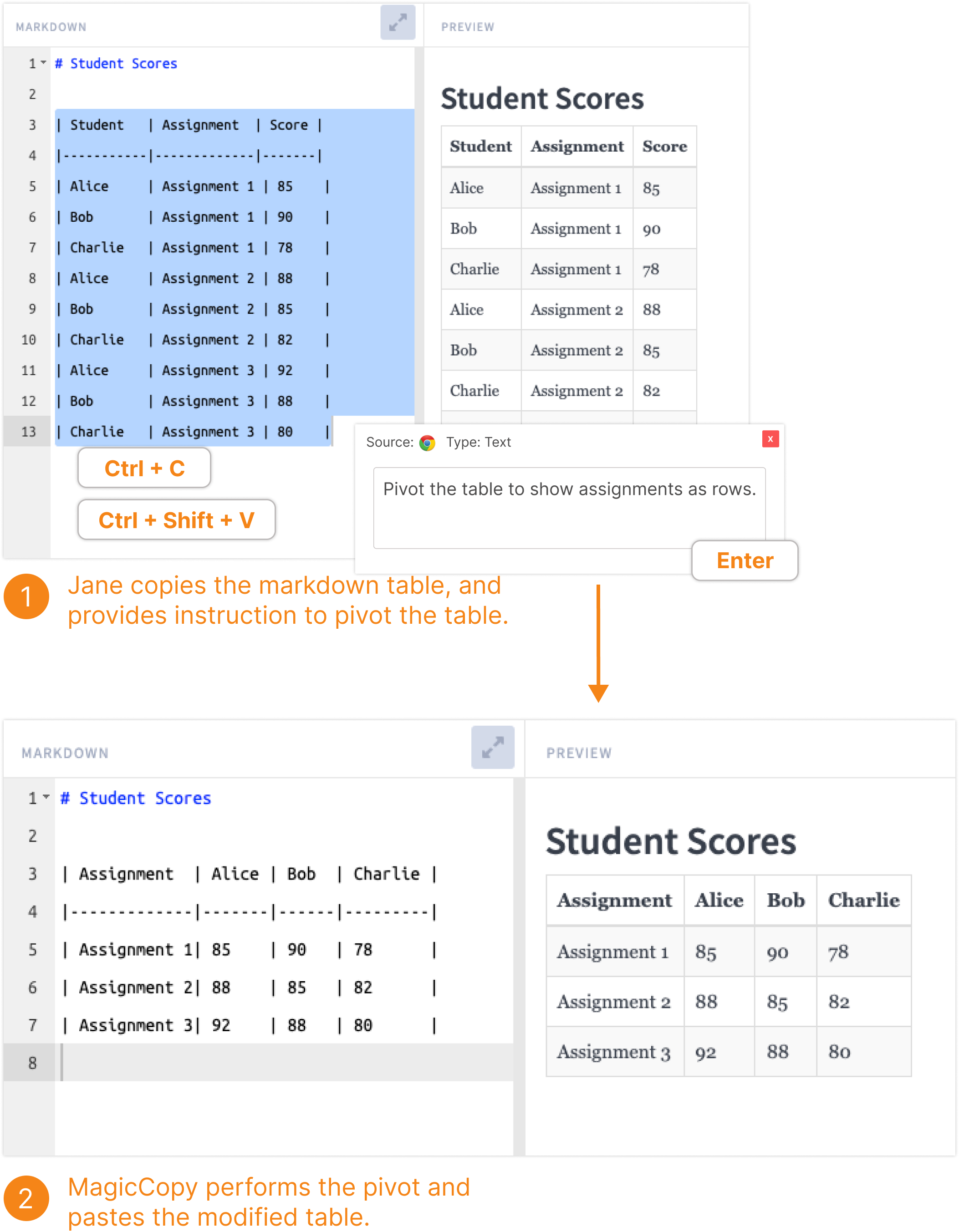}
    \caption{Jane pivots markdown table using \tool}
    \Description{Shows an image of how Jane pivots the markdown table using \tool. She selects and copies the markdown table, then triggers the paste while selecting the copied table, and provides an instruction to pivot the table. \tool automatically parses and pivots the table according to Alice's requirement and pastes the pivoted table back into the markdown editor.}
    \label{fig:scenario4}
\end{figure}

Jane receives an email attachment from her teaching assistant with each student's scores for all the assignments in a Markdown (MD) file exported from the popular note-taking app Obsidian (see Figure~\ref{fig:scenario4}~(1)). The scores are listed in a long format, sorted by assignment. To simplify summary calculations, Jane needs to convert the data to a wide format. As Excel cannot import Markdown directly, she must either input the scores manually or find an online converter to transform the MD file into a CSV. Then, she can pivot the table in Excel to change it from a long to a wide format.


\subsubsection*{\textbf{Scenario 5: Conditional formatting data while copying}}

Susan is a researcher who is testing their model performance over multiple benchmark datasets compared to several baseline models. They run the benchmark using Python Pandas dataframes in a Jupyter notebook. After they finish benchmarking, Alice now has to report their accuracy scores in their research paper as a LaTeX table. She also has to highlight one cell in every row that corresponds to the model with maximum accuracy for the particular benchmark. In addition, to Susan having to manually move her data from a Jupyter Notebook to LaTeX document, she has to manually compute the column with maximum accuracy for each row and format them to bold in the LaTeX representation. 


\subsection{Experience with \tool}

Now let's see how these scenarios are simplified with a simple copy-paste-like experience using \tool, eliminating lot of manual effort. 

\subsubsection*{\textbf{Scenario 1}}


\begin{figure*}[t]
    \centering
    \includegraphics[width=0.7\linewidth]{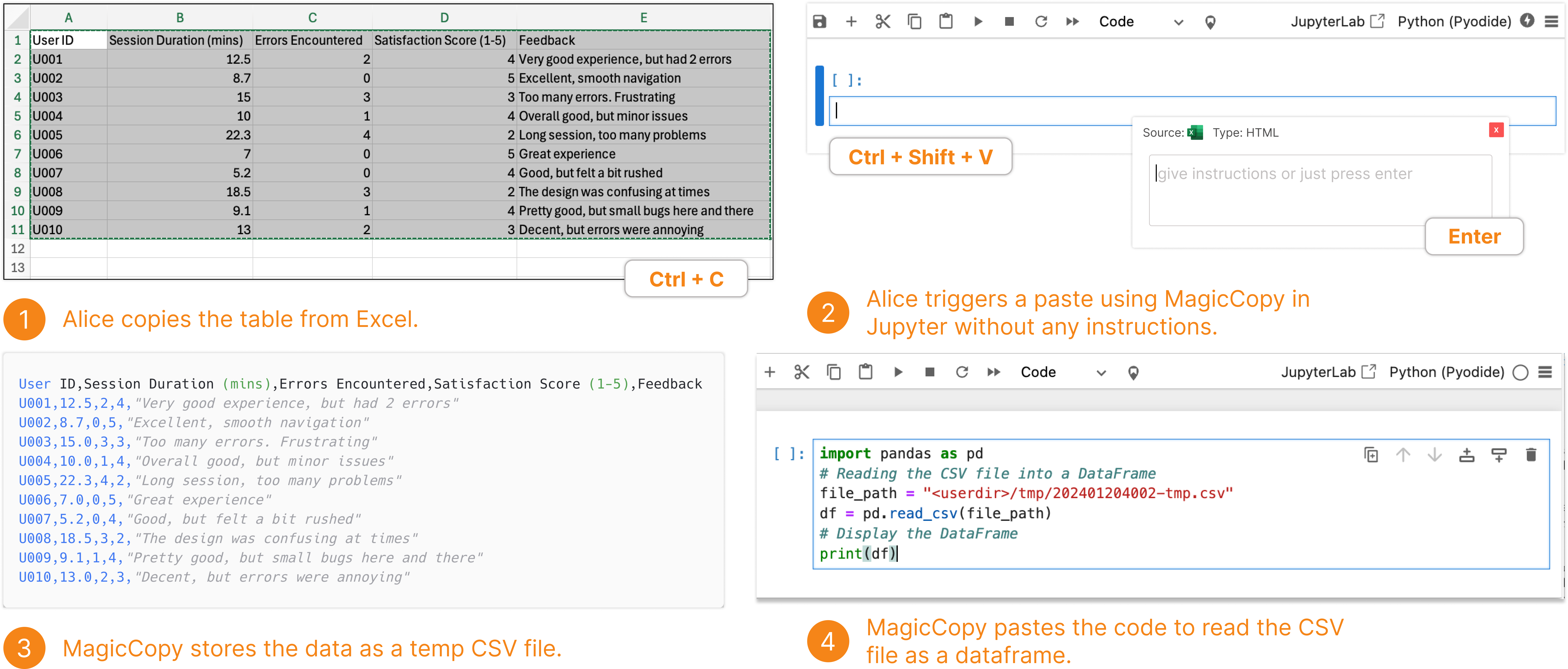}
    \caption{Alice moves the corrected data from Excel to Jupyter Notebook using \tool.}
    \Description{Shows an image of how Alice moves the corrected data from Excel to Jupyter Notebook using \tool. She copies the corrected data from Excel and triggers a paste using \tool without any explicit instructions. MagicCopy automatically stores the copied data as a CSV file and pastes the code to read the file as a dataframe to Jupyter.}
    \label{fig:scenario1}
\end{figure*}

Figure~\ref{fig:scenario1} shows the experience of Alex with \tool. To transfer the corrected data from Excel to Jupyter, Alice simply copies the entire data table from Excel. She then triggers a \texttt{smart paste} in Jupyter Notebook using \tool without any explicit instructions. \tool automatically detects that Alice intends to move data from Excel to a Python code editor in Jupyter. It converts the clipboard's Excel data into a CSV format, saves it as a temporary file, and then automatically pastes the Python code to load the CSV file into a pandas dataframe.

\subsubsection*{\textbf{Scenario 2}}
To transfer the data from Excel to Markdown, Ethan can simply copy the Markdown table, and trigger a \texttt{smart paste} via \tool. The tool automatically converts the Markdown table into RTF format and pastes it into Excel. Now Ethan can go ahead and sort the data, and add computations. similarly, when he is done with the edits, Ethan can simply copy the final Excel table and paste it into his Markdown editor using \tool without any explicit instructions --- the table will be automatically pasted using the Markdown table syntax.

\subsubsection*{\textbf{Scenario 3}}
Similar to the previous scenario, Amy simply copies the PowerPoint table and pastes it into the HTML editor using \tool. This time \tool pasted the table as HTML but without transferring any colors. Amy selects the pasted table and re-triggers the paste. This time providing the instruction to \say{preserve the table colors}. \tool successfully pastes the HTML code for the table with the right colors. 

\subsubsection*{\textbf{Scenario 4}}
Jane's workload is greatly simplified by \tool. Instead of moving the table from markdown to Excel to perform the pivot, she simply copies the markdown table and asks \tool to paste the table with the following instructions: \say{Pivot the table from long to wide format}. \tool performed the pivot operation successfully and directly pasted the data in markdown --- preventing Jane from switching to Excel entirely.


\subsubsection*{\textbf{Scenario 5}}
Writing a LaTex table takes a lot of mundane manual effort. Instead, Susan simply exports the dataframe containing the benchmark results as a CSV file. She then proceeds to copy the contents of the CSV file, and then directly triggers the paste using \tool by giving the instruction \say{bold the highest accuracy values in each row}.

\paragraph{Remarks} As demonstrated, integrating \tool with the copy-paste action provides a unified experience across various applications and data transformation needs. It frees users from switching contexts to an intermediate tool for data prep. Since \tool is context-aware, it can handle complex transformations through simple high-level prompts. In the next section, we explain how \tool generates code based on tracked contexts to support data transformation and formatting.
\section{System Design and Implementation} \label{sec:system}

\begin{figure*}[t]
    \centering
    \includegraphics[width=0.7\linewidth]{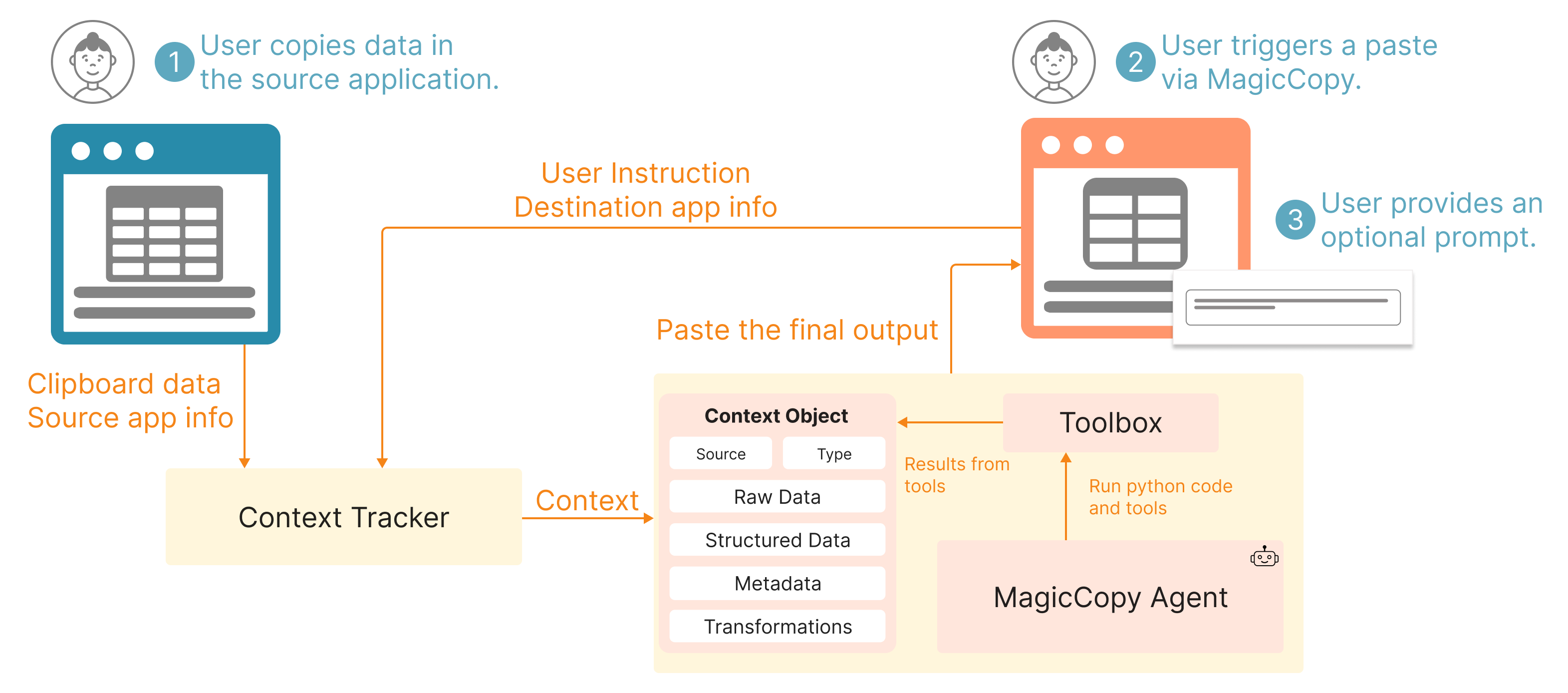}
    \caption{\tool system architecture. The context tracker keeps track of application contexts for copy-paste action and the clipboard context. The \tool agent is responsible for carrying out the transformation and the transfer of data based on the instructions provided by the user. The toolbox contains the tools that will be used by the \tool agent. }
    \Description{Shows an image of \tool's system architecture. The context tracker keeps track of application contexts for copy-paste action and the clipboard context. The \tool agent is responsible for carrying out the transformation and the transfer of data based on the instructions provided by the user. The toolbox contains the tools that will be used by the \tool agent.}
    \label{fig:system}
\end{figure*}

\tool employs three main system design principles: a universal user interface, implicit context tracking, and an action-centric AI agent. The system architecture of \tool is shown in Figure~\ref{fig:system}.

\subsection{Design Considerations: Action-centric instrumental interaction}
Instrumental interaction defines interaction instruments as mediators between users and objects of interest. Similar to how copy-paste acts as an instrument to move simple data between applications, we want \tool to be an instrument to transfer and transform any data between applications effortlessly. For this, we separate the instrument from the information (data) and documents (applications)~\cite{beaudouin2021generative}. We employ two main principles of instrumental interaction for this --- \emph{reification}, \emph{polymorphism}. The combination of user prompts along with the tracking of implicit context \emph{reifies} \tool as an interaction instrument to move any data. The introduction of AI as a mediator between applications to move data allows \emph{polymorphic} use of the tool across a wide variety of data formats and across many applications.

\subsection{Universal User Interface}

Rather than adding yet another standalone app to an ever growing list users must manage, \tool offers a universal, transient user interface --- unobtrusively hidden in the background, ready to engage across any application at a moment’s notice. By implicitly tracking key context, \tool also removes the need to explicitly specify the context of the source and destination applications involved in the data transfer.

\tool is directly attached to the OS clipboard. The user can directly select and copy the data 
in the source app and \sayit{smart paste} via \tool by pressing the keyboard shortcut \texttt{Ctrl + Shift + V} (customizable by the user) in the target app.
%
Since \tool is attached to the OS clipboard, it can support all the formats supported by the clipboard: plain-text, RTF, HTML, and Images. For example, \tool can directly access the image copied from the web and extract data from it, or when the user copies the HTML table, \tool can use the HTML clipboard to infer the structure of the data instead of only relying on the plain-text data (like ChatGPT). 



\subsection{Implicit context tracking}
Successfully moving data between applications requires understanding both the source and destination contexts of the copy-paste action. The \texttt{Context Tracker} in Figure \ref{fig:system} monitors the implicit context of the application from which the data was copied and where the \tool paste was triggered. Using Windows OS hooks, it listens to copy and paste events, capturing details like application name, process ID, and icon. This context is then passed to the \tool agent to help identify the data type and extract structured content and metadata.

For example, when the user copies a table from the web page in Chrome, the context tracker will inform the \tool agent about the application name \say{Chrome}, the tab name, etc. This will help \tool agent to determine the raw data as HTML and appropriately extract the structured data. Later when the user triggers the paste in overleaf, using the context tracker the \tool agent will know it has to transform the paste data in \texttt{LaTex} format.

\subsection{Action-centric AI agent}
\begin{figure*}[t]
    \centering
    \includegraphics[width=\linewidth]{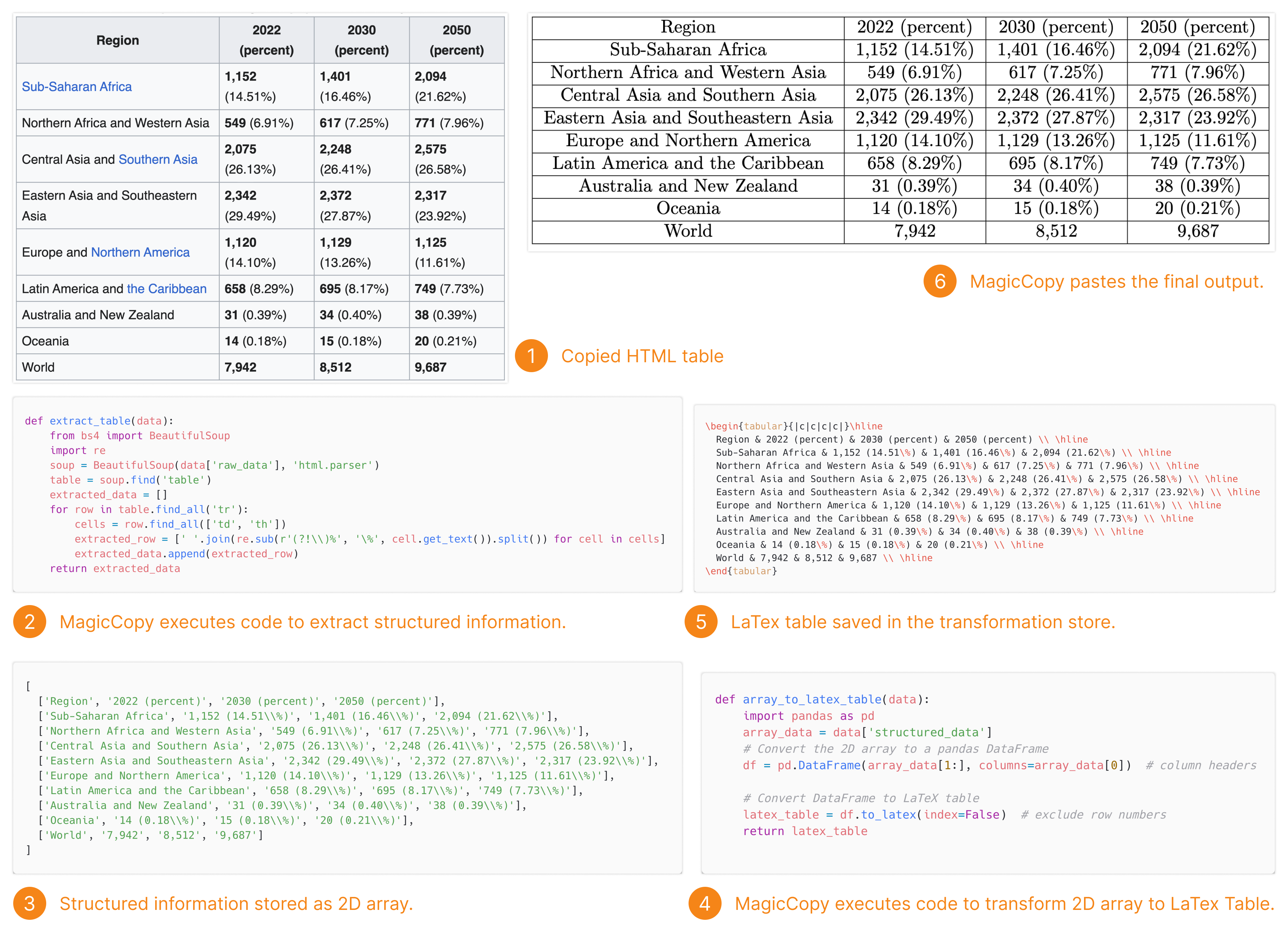}
    \caption{Inner working of \tool --- a simple example. (1) The user is copying an HTML table and pasting it to a LaTex editor. (2) \tool first extracts the structured data from the HTML table by executing a Python function. (3) This structured data is stored as a 2D array. (4) \tool then writes code to transform the structured data into a LaTex table. (5) This transformed LaTex code is stored in the context object. (6) The final LaTex code is then pasted into the target application.}
    \Description{Shows an image of the inner workings of \tool --- a simple example. (1) The user is copying an HTML table and pasting it to a LaTex editor. (2) \tool first extracts the structured data from the HTML table by executing a Python function. (3) This structured data is stored as a 2D array. (4) \tool then writes code to transform the structured data into a LaTex table. (5) This transformed LaTex code is stored in the context object. (6) The final LaTex code is then pasted into the target application.}
    \label{fig:sys-ex1}
\end{figure*}

Starting with the clipboard data, user instruction, and source and target app information, \tool performs a series of transformations on the data using an AI agent to convert it to the format needed for the target app (as shown in \autoref{fig:sys-ex1}). \tool internally organizes the data context in multiple formats---raw data, structured data, metadata, and transformations (see \autoref{fig:system}) inside its \texttt{Context Object}, in order to support the various data-format transformations needed for moving data between multiple apps. Raw data is obtained directly from the OS clipboard and it can be of four types: Text, HTML, RTF, or Image.  The rest of the formats are derived by \tool's AI agent as needed. 

\paragraph{Structured data.} Since the raw data can be in many different formats (HTML, CSV, space separated, string), \tool first parses the raw data from clipboard and extracts the data contents into a simpler format such as 2D array (\autoref{fig:sys-ex1}(3)). The format of this structured data is dynamically chosen by the AI agent based on the raw data as well as the user instruction. For example, for another data transfer task in which multiple tables are copied in the source app, \tool used an array of dictionaries as the structured format (\texttt{[\{'headers':[...], 'rows':[...]\}, ... ]}). 

\paragraph{Metadata.} \texttt{Metadata} is a key-value store that can be used to capture other auxiliary information from the raw data that is not captured by the structured data such as colors and font formats. Again, the AI agent is free to choose the keys and values (which itself could be another array or dictionary) to add to the metadata.

\paragraph{Transformations.} \texttt{Transformations} is another key-value store that records all the transformations performed by \tool on raw data, structured data, and metadata. These include transformations to add/remove columns, to add conditional formatting, or to transform to target app's data representation, etc. \autoref{fig:sys-ex1}(5) shows the structured data in (3) transformed as a string representing the corresponding LaTex table.

\paragraph{Code generation rather than data generation.}
To generate the above data representations and transformations, \tool's AI agent generates Python code to programmatically extract and transform the clipboard data, rather than directly generating the new data. For example,  Figure~\ref{fig:sys-ex1} (2) shows the code generated by the AI agent to parse the HTML table in (1) into a structured 2D Python dataframe in (3) using the \texttt{BeautifulSoup} package. By generating and executing code to generate data transformations, \tool can operate on tables that may not fit fully into the AI model's context window as well as reduces the chance of the model hallucinating the data it generates. 

\paragraph{AI agent's design using composable tools.} Moving data from one format to another often involves a series of programmatic transformations on the different data representations in \tool's \texttt{context object}. These steps are context-dependent, requiring the AI agent to dynamically plan and execute a set of steps. To support this, we provide the agent with a small set of composable tools used alongside OpenAI's assistant API~\cite{openai-assistant}.
These tools give the agent agency to choose specific parts of \tool's context to operate on, execute generated code, interact with the file system selectively, and paste the transformed data into the destination. The \tool agent understands the internal clipboard representation (Figure~\ref{fig:system}) and can use its tools to inspect specific parts of the data—such as sampling structured content or accessing stored values in metadata or transformation object stores. The complete list of tools available to \tool is attached with the supplementary materials.


\paragraph{Beyond copy-paste: interacting with apps via plugins}
Not all data transfer tasks can be simply completed through copy-paste actions, especially for complex applications like Microsoft Excel. Recreating complex properties like setting datatypes or specifying column width can only be done by interacting with the Excel GUI. This requires applications to have deeper integration with \tool. 

To support deeper integration, \tool allows external applications to provide custom APIs by subscribing via web sockets. These APIs enable the \tool agent to access richer context around the paste action and perform tasks beyond the limits of standard copy-paste. When available, the agent is instructed to prefer using these APIs, falling back to copy-paste only when APIs are not available for a given application.

As a proof of concept, we implemented a plugin for Microsoft Excel, which gives \tool complete access to the Office Excel Javascript APIs. Note: The data transfer to Excel works even without this plugin using just the copy-paste interface. However, the users can perform more complex tasks like setting datatypes, pasting the data on a new sheet, etc. using the plugin. 


Note that all tasks in the user study can be performed with or without the Excel plugin. The plugin serves as a proof of concept to demonstrate the extensibility of the \tool implementation.

\subsection{Implementation}
Currently, \tool is implemented as a Windows application (due to the reliance on Windows native APIs for context tracking). 
\paragraph{Context Tracker} We used a combination of Windows native API hooks and the \texttt{pywin32} Python package to track application information for both the source of copied data and the destination of pasted data. Specifically, we collect the application's name, icon, and PID. To access clipboard data, we use Electron clipboard APIs~\cite{electron-clipboard}.
\paragraph{\tool Agent and GUI} We use OpenAI's assistant APIs~\cite{openai-assistant} with the GPT-4o~\cite{gpt4o} model for the \tool agent. GPT-4o is a high-performance, multi-modal model that excels at accurate function calling while maintaining fast response times. Although we have also tested \tool with GPT-4o-mini~\cite{gpt4o-mini}, its faster responses often come at the cost of code reliability, leading to a longer overall runtime for task completion. We expect smaller models to improve over time, making them more viable for practical use in the future.
The \tool agent is implemented as a Python backend, with a front-end GUI built using Electron and React. Communication between the GUI, Python backend, and subscribing application plugins is handled via web sockets. These plugins can provide custom APIs to extend the agent’s capabilities.
The agent also operates within a restricted Python sandbox, which grants limited permissions to executed code. It supports widely-used packages like pandas, numpy, matplotlib, BeautifulSoup, and more—this set can be customized by installing additional packages in the environment. We automatically detect static syntax errors or deviations from the expected syntax and pass error messages to the model for correction attempts.
In the case of runtime errors during code execution, the error message is again passed to the agent, which may retry up to three times. If all attempts fail, the process is gracefully terminated and an error message is displayed to the user.
\vspace{-0.8em}
\section{Study Design and Rationales}


We aim to study how \tool facilitates data movement across different applications and opens up new interaction possibilities. We ground the objective into the following research questions:

\begin{itemize}[leftmargin=6pt]
    \item \textbf{RQ1} (User experience) What is the perceived user experience of using \tool and how do users compare it with traditional copy/paste?
    \item \textbf{RQ2} (Expectations and Strategies) What are users' mental models and expectations about \tool, and how do they affect users' strategies when working with \tool?
    \item \textbf{RQ3} (Application scenarios) Where do users envision bringing \tool to their workflow?
\end{itemize}

\noindent With these questions in mind, we designed a study structured in two parts, with \tool as a technology probe ~\footnote{Design probes in HCI are adapted from cultural probes~\cite{gaver1999design}, designed objects that aim to promote user engagement in the design process~\cite{boehner2007hci}.}~\cite{graham2008probes, boehner2007hci, hutchinson2003technology}. In the first part, we ask participants to use \tool to move datasets assigned by us across three pairs of applications, with the goal of learning users' experiences and strategies of using \tool to move data across applications. Then, in the second part, to learn where users could apply \tool to their workflow and how they perceive it compared to processes based on traditional copy-paste, we conduct an open exploration activity, asking participants to apply \tool to any set of applications and datasets of their choosing that suit their workflow.

\subsubsection*{\textbf{Study setup}}  We recruited 16 participants (9 female, 7 male) through the mailing list of a research university. Participants reported a variety of experiences with Generative AI (GenAI) tools like ChatGPT, Gemini, Dall-E, Claude, GitHub Copilot, Adobe Firefly, and more. Of the 16 participants, 9 participants use GenAI tools almost daily, 3 participants use them multiple times per week, 2 participants use them a few times a year, and the remaining 2 participants use them very rarely. Of the 16 participants, 12 are between 18-25 years old, 3 are between 26-30, and one is between 31-45.

We designed the study session to take 1 hour. After introduction and obtaining consent (5 minutes), participants complete a 5-minute tutorial to learn the basics of \tool. Then, participants are given 15 minutes to complete the first part of the study (specific tasks) to complete 2 designed data transfer tasks, and another 20 minutes for the second part (open exploration). After each part, we ask participants to fill out the post-study survey and conduct an interview with the participants. 
We recorded the audio and screencast for each participant and asked them to think aloud during the study. We present the detailed designs. Participants received \$25 as compensation for their participation.

\subsubsection*{\textbf{Study part 1: transferring data across assigned applications}} \label{ssec::tasks}

In this part of the study, we have a pool of three datasets from the Web, with each assigned three distinct sub-tasks (see Table~\ref{tab:study-tasks}). We also have a pool of three applications, each requiring a different representation --- Excel, LaTex, and Markdown editor. For the purpose of counterbalancing, we have prepared a copy of these datasets following their original format in Excel, LaTex, and Markdown.

Each participant is given two datasets ($D_1, D_2$) and two applications ($A_1, A_2$). For the first dataset $D_1$, the participant performs the sub-tasks between the web and the first application $A_1$. For the second dataset $D_2$, the participant performs the sub-tasks between the first application ($A_1$) and the second application ($A_2$).
In total, the participants must complete 6 data transfer tasks (2 datasets x 3 sub-tasks). The order of the datasets and applications is counterbalanced.

To best probe participants' feedback in different data transfer scenarios, we choose the following three datasets from the Web, each features a different formatting style and structural presentation: 
\begin{itemize}[leftmargin=*]
\item Dataset 1 (Generic Ballot Polls, 2024 USA elections) This dataset comes from a data journalism website \texttt{FiveThirtyEight} showing the generic ballot polls for the 2024 USA elections~\cite{2024genericballot}. In this webpage, the data is split into multiple HTML tables, with complex background coloring based on values.
\item Dataset 2 (Presidential Election Polls, 2024 USA elections): This dataset also from \texttt{FiveThirtyEight} shows the presidential election polls for the 2024 US elections~\cite{2024presential}. The data is formatted in a single HTML table with complex background coloring based on values. 
\item Dataset 3 (2020 Tokyo Olympic Medals Table: Artistic Gymnastics; Japan) This dataset from the official Olympics website shows the medal table for the 2020 Tokyo Olympics~\cite{2020olympics}. The data is represented in a single HTML table with mixed text and image content. In each row, the last column is composed of three sub-rows with three medal types that are represented as icons.
\end{itemize}

\autoref{tab:study-tasks} shows the dataset and task pool from which two datasets are chosen for each participant. These tasks cover five types of transformation and reformatting tasks users often encounter in practice: 1) removal of columns, 2) addition of columns through computations, 3) merging two columns, 4) splitting a column into multiple columns, and 5) conditional highlighting.


\begin{table}[h]
    \centering
    \resizebox{\columnwidth}{!}{%
    \begin{tabular}{|c|l|}\hline
        \textbf{Dataset} &  \textbf{Tasks}  \\\hline
        Ballot Polls  & \begin{minipage}{3.0in}
    \vskip 4pt
    \begin{enumerate}[leftmargin=*]
        \item Copy the table without any transformations. 
        \item Paste the table without the fourth and fifth columns.
        \item Paste the table and merge the second and third columns.
    \end{enumerate}
   \vskip 4pt
 \end{minipage}\\
 Presidential Election  & \begin{minipage}{3.0in}
    \vskip 4pt
    \begin{enumerate}[leftmargin=*]
        \item Copy the table without any transformations.
        \item Paste the table without the last two columns.
        \item Add a column to show the difference in the polling percentage.
    \end{enumerate}
   \vskip 4pt
 \end{minipage}\\
 Olympic Medals & \begin{minipage}{3.0in}
    \vskip 4pt
    \begin{enumerate}[leftmargin=*]
        \item Copy the table without any transformations.
        \item Split the \say{Medals} column into three by medal type.
        \item Highlight athletes with at least one gold medal. 
    \end{enumerate}
   \vskip 4pt
 \end{minipage}\\\hline
    \end{tabular}%
    }
    \caption{Data moving tasks in the first part of the user study.}
    \Description{Shows a table containing three datasets: Ballot polls, Presidential election, and Olympic medals. Each of these datasets is assigned three distinct tasks for the user to complete.}
    \label{tab:study-tasks}
\end{table}

We prepare these tables following their original formats in Excel, Latex, and Markdown Editor so that different participants can work with different source and target applications for each dataset. We counterbalance the order of the tasks and the source-destination applications across participants. These tasks are challenging data transfer tasks that cannot be completed with regular copy-paste and/or ChatGPT as an intermediate tool due to the complexity of structure and formats needing copying. Through our study, we can elicit participants' reflections about how \tool can enable them to perform these challenging tasks and augment their standard data transfer practices.

During this part of the study, participants are allowed to retry as many times as they choose should they encounter challenges using \tool. A sub-task is considered failed if the participant gives up the problem or runs out of time.  After participants completed the tasks, participants completed a post-task survey containing questions from the System Usability Scale, and NASA's Task Load Index (See supplementary materials), and provided answers to a semi-structured interview that collected qualitative feedback about \tool. Upon completion, participants move to the next part of the study, where we ask participants to explore potential applications of \tool in their workflow.  
We measured the success/failure for each sub-task the participant had to perform and recorded all the natural language commands the user provided to \tool. We also record all interview responses for later analysis.

\subsubsection*{\textbf{Study part 2: open exploration}}

In the open exploration part of the study, we gave participants free reign with \tool, letting them use it on any set of applications and datasets of their choosing. We encouraged participants to use the dataset and test tasks that were closer to their real-world use. The study conductor was available to help the participant when needed. This part of the study closely resembled contextual inquiry studies~\cite{holtzblatt1997contextual} where the conductor sought to observe and understand how users applied \tool. The conductor was allowed to interject to ask clarifying questions and help with any technical difficulties faced by the participants. 

Upon finishing the exploration,  we asked participants to self-evaluate if they ``succeeded'' in each data transfer task they explored and elaborate on their experience. We also asked participants to elaborate on their experiences with \tool and how they see \tool fitting into their daily workflow. We later used these answers in our study analysis.

\subsubsection*{\textbf{Analysis}} After all the study sessions, we analyzed participants' usage patterns and experiences with \tool based on their success measure, survey results, telemetry data, and interview responses. The first author performed open coding on the participants’ responses to interview questions and the audio transcripts to identify themes and then discussed with co-authors to refine the themes over multiple sessions. These themes are used to explain the qualitative results. 

\section{Study Results}

In this section, we first present participants' task completion metrics and report their experiences of using \tool to transfer data. Then, we presents participants' strategies working with \tool, and scenarios they expect \tool to help them with their personal work.

\subsection{Task completion and user experiences} All the participants successfully completed all the specific tasks in the first section of the study. Each participant took an average of 1.125 ($\sigma=0.33$) attempts to complete each sub-task with \tool. There were 12 instances (out of 96) in which the participant had to retry the task, and of those they only had to retry once. Five of these instances were due to a partial downtime in OpenAI's APIs that resulted in the request timing out, and the rest (7) of these instances were due to model error.
System usability scores (See attached figure with supplementary materials) show that all but one participant reported that it was easy to perform the tasks with \tool. In the post-task survey, where participants self-reported cognitive load, 87.5\% indicated that they felt less \emph{mental demand} and \emph{hurry} when using \textit{\tool}. Additionally, 81.25\% reported that \textit{\tool} reduced the \emph{effort} required to complete the task. All participants expressed a strong sense of \emph{success} and reported very low levels of \emph{annoyance}.

We organize participants' experiences completing the tasks in the following three themes.

\subsubsection*{\textbf{Theme 1: \tool enables more streamlined interactions across applications by reducing manual efforts}}

Thirteen participants (P2-3, P5-14, P16) mentioned that \tool can streamline the workflow by removing manual intermediate steps in the data-moving process. When we asked them to reflect on how their experiences with \tool stack up against their past experiences involving traditional copy-paste and GenAI, participants provided valuable insights. 

With traditional copy-paste, transferring data with complex formats can fail, and participants often need a significant amount of manual effort typing, exporting/importing with intermediate files (e.g., CSV), and then reformatting to ensure that all data and its semantic structure are transferred over, and its rendered at its destination properly. For example, P2 mentioned \sayit{[regular] copy-paste never works, usually, I just type it myself}, and P7 said \sayit{I will admit that I have tried copying a table from Wikipedia or the Internet and pasting that onto Excel. But the format has always been funky and messy}. Although GenAI tools today can make data conversions from one representation to another a straightforward task, it still requires the participant to juggle between three or more tools to move the data with these intermediate tools.  
For example, P7 mentioned \sayit{I think it [\tool] definitely removes that layer of moving to another app like ChatGPT to do it. It [ChatGPT] will give me a code, and then I’ll have to paste that code into an editor, run it, copy the output, and paste it into the final app. This just kind of cuts to the chase -- copy-paste, and done}.


Whereas when using \tool, because participants can provide transformation and formatting instructions when copying data, they find that it saves effort and time. For example, P6 said \sayit{The biggest takeaway from today is that I saved a lot of time that I would have otherwise spent manually editing}. P13 said \sayit{This was fast. I have spent a lot of time cleaning and converting data. This really improves the workflow}. 


\subsubsection*{\textbf{Theme 2: A \tool's key advantage is its ability to dynamically adapt paste formats based on contexts and user instructions}}

Participants acknowledged that \tool's key strength is its ability to track source and destination applications and then format data based on user instructions.
Eight participants (P1, P2, P3, P8, P10, P11, P12, P13) explicitly highlighted that they liked the adaptability of \tool to dynamic and multi-modal content. P3 summarized by saying \sayit{I'm pretty impressed by its compatibility with so many different platforms, as well as different data formats like unstructured text, tables, images}  There are many instances where participants exploited the context tracking of \tool. For example, when P10 pasted the tabular data from the web into an R file editor without any additional prompt, \tool automatically converted the HTML table into a CSV file along with the required filtering and transformation. Then \tool saved the file to a temporary directory and pasted the code to read the CSV file as an R dataframe into the editor. P10 was impressed by this when they said \sayit{This tool is instantly adaptable with the data I have and what I say. It just pasted the code in R directly. Normally, what I find the most time-consuming is the code part of bringing the data}. Similarly, when P1 wanted to extract the tabular data from a malformed CSV file, \tool automatically generated code to remove the malformed parts before converting it to markdown format. Ten participants (P1, P2, P3, P4, P6, P8, P9, P10, P13, P15) explicitly mentioned that this dynamic adaptation aspect is useful to solve tasks in their workflows. 
As we will describe in the next section, \tool's context awareness affects users' prompting style when providing instructions for data transformation.

\subsubsection*{\textbf{Theme 3: The copy-and-paste metaphor made \tool intuitive and accessible. It also made it inappropriate for large datasets.}}
Twelve participants (P2, P4-10, P12, P13, P14, P15) explicitly mentioned that they found \tool's copy-paste interaction model intuitive and easy to use to perform the tasks. Participants highlighted the simplicity of the action -- P5 said \sayit{I think the copy-paste interaction makes this unique. Makes it simple to use}. P12 said, \sayit{I like the fact that it's quick and dirty -- just copy and paste it how you want. It's really helpful}.
Using copy-paste as an underlying interaction metaphor allowed participants to easily access and benefit from \tool's capabilities. During the study, seven participants (P2, P3, P6, P7, P9, P10, P13) explicitly made reference to this. 
P9 said \sayit{Very convenient in terms of opening and telling it what to do! I enjoy that it \say{follows} me where I work and is accessible via keybinding}. Also, P5 said \sayit{I think it's cool that this is kind of like ChatGPT in your hands, where you can give it some prompt and paste it the way you want.}

Although simple, five participants (P3, P11, P12, P13, P15) noted that copy-paste is not a great way to move large amounts of data. P13 expressed this by saying \sayit{I never thought about copy-pasting between apps. I think one reason is that the data that I often work with are very large. So like, I kind of just assumed, I wouldn't be able to copy-paste}. P15 even assumed limitations to the clipboard when they said \sayit{Since the tool uses the clipboard -- that sort of limits it to a smaller data set}.

\subsubsection*{\textbf{Theme 4: \tool's response delay could be improved, but it still saves time compared to traditional approaches}}
Since \tool uses LMM models to parse and transform the data, \tool responses are modulated by the LMM's response times. This is a major difference from the traditional copy-pasting action, which is almost instant. This delay is the combination of data sampling cost, LMM's code generation time, and the cost of executing Python code to extract structured data. The delay can be around $\approx30-40$ seconds between when the user triggered the paste and the actual pasting of the data. Ten participants (P2-3, P7-13, P15) said they would like a reduced delay in the time taken to get results. Two participants (P10, P12) conveyed a strong dislike of this delay: P12 said \sayit{It takes some time. Immediate feedback is important to me} -- as they compared to how instant regular copy-paste is. 

Despite the delay, 12 of the 16 participants (P2, P4, P6-10, P12-16), including those who were apprehensive of \tool's delayed response, acknowledged the time saved by the tool (compared to their manual processing efforts) far outweighs \tool's response delay. P10 said \sayit{it nets me time like I didn't have to spend that time manually looking up things and changing things}. P13 said \sayit{But the advantage is if I had to manually copy and format it would take way longer than this}. 
Seven participants (P1, P4, P6, P10, P13-15) mentioned they would roughly assess the time required to manually complete a task versus using \tool to do it, and then decide accordingly. They further elaborated that deciding when to use \tool depends not only on the task but also on the application. P10 said \sayit{I think deleting two columns in Excel wouldn't take time to do it myself. But for Markdown, I would have to manually find each value and delete it, so I would just use the tool [\tool] here}. Similarly, P13 said \sayit{For table operations, I would do it myself in Excel. But for tools like LaTeX or Markdown, I will have to use multi-cursor. So it is easier with the tool [\tool]}.


\subsection{Expectations and strategies: How did participants interact with \tool?}

In the post-task survey, all participants mentioned that they could easily communicate their goals to the system. All but one participant mentioned that they could easily plan and manage their tasks using \tool, and maintain control of their creative process. This was also observed during the study, where 11 participants explicitly (P3-6, P8, P10, P12-16) mentioned that \tool understood their intent well. In our analysis of participants' prompts and interview results, we noticed different interaction styles (especially prompting and verification styles) used by the participants, depending on their prior GenAI experience, the mental model of \tool, and their background.

\subsubsection*{\textbf{Observation 1: Participants prompt \tool differently based on their experiences}}

We observed two distinct prompting styles among participants, likely influenced by their prior experience with GenAI tools. Eleven participants always gave a short and succinct prompt, while five participants (P3, P4, P5, P10, P14) gave long and specific prompts to ensure there was no ambiguity with the intent. 
Some participants even intentionally provided vague prompts during the open exploration to see if \tool would respond correctly, and after their success, they enjoyed communicating their intent with shorter prompts. For example, P15 gave the prompt \sayit{Create a pivot table of class by number of survivors} to see if it correctly recognized the columns \say{Pclass} and \say{survived} in the Kaggle Titanic dataset to aggregate while pivoting. \tool successfully created the pivot table using both columns. Similarly, P1 gave the prompt \sayit{Remove the 8th and the 9th column}, when pasting a table with only seven columns. Even in this instance, \tool pasted the table correctly without removing any columns.  

Participants also did not have to provide any information about the format due to \tool's inherent context tracking -- which enabled the users to provide shorter and more succinct prompts. For example, P15 said \sayit{This is so cool. I didn’t expect it to know what to do. Generally, I will have to give the AI a lot of information. This is pretty helpful and unexpected} -- referring to how they did not have to provide the format information for the destination application. When asked why they prefer longer prompts, P14 said \sayit{In my experience with generative AI in general, the more specific you are the more likely it'll be successful}. 



\subsubsection*{\textbf{Observation 2: Verification is essential for participants to work with \tool.}}

In the post-task survey, $87.5\%$ of participants reported being happy with the system’s output, and $75\%$ felt confident that others would also rate it highly. Additionally, $75\%$ said they felt a strong sense of ownership over the creative outcome when using \tool. These high ratings align with the tool’s strong task success rate.

In general, participants find verification important when working with \tool. Six participants noted that their trust in \tool was shaped by past experiences with AI tools. As P6 put it, \sayit{I would double check. Just because with AI tools that I have used in the past there, it’s not always perfect}. This skepticism towards AI prompted participants to actively verify \tool’s output—indeed, all participants manually checked the results against the original data to ensure correctness.

When participants used \tool to move large datasets or perform complex transformations like aggregation or pivoting (as in the second part of the study), verifying results became difficult and impacted their trust. Ten participants (P1, P2, P4, P5, P8, P9, P11, P13, P15, P16) noted that verifying outputs for such tasks would require significant effort—making them hesitant to use \tool in these cases. As P9 noted, \sayit{As we get to larger datasets, obviously, I cannot look through every single row. I’m probably not going to do that}. P16 similarly remarked, \sayit{I would just be concerned that there wouldn’t be an easy way to detect errors}. This concern was reflected in trust scores: only $56\%$ of participants said they would trust the system’s output. Despite these concerns, participants readily used \tool for a variety of complex tasks—often beyond the intended scope in both data and application—and succeeded in many cases, as detailed in section~\ref{ssec::oe}.

\subsubsection*{\textbf{Observation 3: How participants use \tool is influenced by their understanding or perception about how it works..}}

We observed that participants often used \tool in ways shaped by their mental models, which sometimes differed from how we designed it. For instance, trust in \tool’s output was influenced by how participants thought it worked. After we explained \tool’s workflow --- specifically that it uses code to extract transformed data --- six participants (P1, P8, P10, P12, P13, P15) said this increased their trust. As P12 noted, \sayit{If it is using code, its thinking process is more rigorous. Now I trust the values a bit more}. Additionally, eight participants (P1, P5, P8, P10, P12, P14, P15, P16) said they would like access to intermediate code and outputs for understanding, debugging, or reuse.

We saw similar mismatches in expectations around follow-up commands. While most had no issue using them, five participants (P4, P5, P8, P10, P11) were unsure if \tool retained clipboard context\footnote{\tool automatically preserves clipboard context for follow-up commands.}, leading them to repeatedly copy the data before issuing follow-ups. Participants also had differing expectations about formatting. Six (P15, P7, P8, P4, P11, P13) assumed formatting would be preserved across applications. Others focused more on raw data: three (P1, P2, P5) wanted the ability to set a default format preference and override it with a prompt when needed.

\subsection{Application scenarios: How do participants expect to use \tool in their workflow?} \label{ssec::oe}

\begin{table*}[t!]
\centering
\begin{tabular}{|p{0.1\linewidth}|p{0.53\linewidth}|p{0.33\linewidth}|}
\hline
\textbf{PID} & \textbf{Explorations}                                                                                                                                                                                                           & \textbf{Dataset}                                                  \\ \hline
P1                   & \textcolor{tsuccess}{Parsing and cleaning malformed CSV files, de-duplicating tables.}                                                                                                                                                                            & Fake simulation dataset to resemble work data                                \\ \hline
P2                   & \textcolor{tsuccess}{Computed columns, conditional formatting, pie chart (LaTex)}                                                                                                                                                          & World population data from web                                    \\ \hline
P3                   & \textcolor{tsuccess}{Image to tabular data, table to Python dataframe, unstructured text to Table.}                                                                                                                                                            & World population data from Wikipedia, Sample unstructured data from web \\ \hline
P4                   & \textcolor{tsuccess}{Table sorting, computed columns, Image to Table}                                                                                                                                                                                 & Yahoo Finance Stock data                                          \\ \hline
P5                   & \textcolor{tsuccess}{Table cleaning, computed columns, table to dataframe, pie chart (LaTex)}                                                                                                                                                                   & GDP dataset from web                                              \\ \hline
P6                   & \textcolor{tsuccess}{Summary statistics (mean, median), cleaning missing values, type annotation, plotting in Python}                                                                                                                                                          & Sensor data from work                                             \\ \hline
P7                   & \textcolor{tsuccess}{Image to table data, conditional colored highlighting, unstructured logs to structured tables}                                                                                                             & Sample table image from web, Sample logs from web (IBM blog)      \\ \hline
P8                   & \textcolor{tsuccess}{Answering questions based on table aggregation, Filtering gene expression data based on rules and world knowledge.}                                                                              & Gene expression data from web, World population data from Wikipedia              \\ \hline
P9                   & \makecell[{{p{\linewidth}}}]{\textcolor{tsuccess}{Sorting, computed columns.} \\ \textcolor{tfail}{Failed: Data extraction from URL.}}                                                                                                                                                  & Kaggle Mushroom binary classification dataset                     \\ \hline
P10                  & \textcolor{tsuccess}{Cleaning table data containing images, icons, links, and citations; computed columns in dataframe, reshaping dataframes.}                                                                                                 & Stadium dataset                                                   \\ \hline
P11                  & \makecell[{{p{\linewidth}}}]{\textcolor{tsuccess}{Unstructured data to table from PDF using both regular copy and image. cleaning specific special characters.} \\ \textcolor{tfail}{Failed: Image to table missed some rows}}                                                        & Custom unstructured data from a research PDF.                     \\ \hline
P12                  & \makecell[{{p{\linewidth}}}]{\textcolor{tsuccess}{Computed columns, filtering.} \\ \textcolor{tfail}{Failed: Error parsing chinese characters using the tool.}}                                                                           & World population data, Chinese Rime Table images from web.        \\ \hline
P13                  & \textcolor{tsuccess}{Table data to python string to dataframe. Multiple tables to multi-indexed dataframes, Image to table data.}                                                                                           & Generic US ballot polls (task dataset)                            \\ \hline
P14                  & \textcolor{tsuccess}{Testing multiple operations in one prompt, creating custom composite functions based on a combination of test scores.}                                                                                                             & Stadium dataset, Fake behavioral test scores dataset.                         \\ \hline
P15                  & \textcolor{tsuccess}{Data cleaning, calculated columns, pivot tables}                                                                                                                                                                                 & Kaggle Titanic dataset                                            \\ \hline
P16                  & \makecell[{{p{\linewidth}}}]{\textcolor{tsuccess}{Answering questions based on computations on the table, Image to table data. Unstructured table data to structured data.} \\ \textcolor{tfail}{Failed: An image with unusual formatting failed to convert to a table.}} & Table data from work.                             \\ \hline
\end{tabular}
\caption{Scenarios explored by participants in the open exploration.}
\Description{Shows a table containing the scenarios explored by each of the 16 participants in the open exploration.}
\label{tab:oe}
\end{table*}

In the open exploration stage, we encouraged the participants to move data between any applications they wanted. Participants used a variety of applications (Excel, Overleaf, Markdown editors, Windows Notepad, Python notebooks, Online R editors, Visual Studio, and Visual Studio Code) and performed a wide range of tasks (data cleaning, plotting, etc.) across many types of data (HTML tables, Excel tables / Markdown / Latex, Unstructured text from PDFs and webpages, images containing both table and unstructured text, and CSV/TSV files). Despite not having tested \tool on these apps and data types, participants were successful in most of their explorations. 

Table~\ref{tab:oe} shows all scenarios explored by participants, where we highlight the dataset, application, and tasks they attempted (we label the tasks they succeeded in \textcolor{tsuccess}{blue} and the ones with which they had problems in \textcolor{tfail}{red}). We categorize their exploration into the following six scenarios based on the intended \emph{formatting and transformation to be performed} by the participants along with the data transfer.


\subsubsection*{\textbf{Data transformation and conditional formatting}}
After exploring data transformation and formatting in the first part of the user study, most participants would like to experiment with these transformations again, but with their own datasets as well as with more complex operations. For instance, P13 recreated the conditional highlighting using colors based on the values of the cell from a plain table.  P14 added composite columns that spanned computations across multiple columns. P9 and P15 added computed columns on a Kaggle dataset that will be useful in their prediction tasks. P15 even asked \tool to perform pivot tables to understand \say{survival rate} compared to \say{gender} and \say{ticket class}. 

\subsubsection*{\textbf{Data Cleaning and Filtering}}
Several participants explored the potential of using \tool for data cleaning, especially when they only need a small tidying up of the data before using it for analysis. For example, P6 presented a sensor dataset they worked on with many missing values spanning across 36 columns, and they experimented using \tool to try three ways to impute the data -- empty, 0, and mean. All three attempts succeeded, where \tool generated the correct imputation code based on their instructions. P1 tested \tool with a malformed CSV data (created on their own) that resembles a simulation dataset used in their work. They asked \tool to extract data on relevant rows and also de-duplicate data based on a combination of columns to the Python editor. P10 applied \tool to a Wikipedia dataset containing images, flag icons, citations, and links to perform a filtering operation to obtain only the relevant columns with the wanted data.

\subsubsection*{\textbf{Copying data to be used in programming environments}}
Many participants explored using \tool to assist in programming, where they moved data from Excel or Web pages to Python pandas data frames to reduce manual efforts in importing and exporting the data. Participants P1, P2, P5, P6, P10, and P13 explored converting tabular data to dataframes. P6 used \tool to convert CSV data to a dataframe and were pleased with the fact that \tool could appropriately identify datatypes of each column like float, and string, that they otherwise would need to process manually. P13 used \tool to merge and move multiple tables into a dataframe while maintaining the cluster of rows within each table using multi-indexing. 

We had an interesting observation with P1 where they used \tool to generate code. P1 was using \tool to clean malformed CSV data and convert it to dataframes within a Python editor. The tool automatically cleaned the CSV data, removing the malformed lines, saved the final output as a temporary CSV file, and then pasted the Python code to read the CSV file into a dataframe. However, P1 was curious to understand how \tool achieved this. Therefore, they pasted the data with a new prompt: \say{Paste the data as a string variable without any change and then convert it to a dataframe}. This forced \tool to do the data cleaning via the pasted code. This was an interesting use case that we did not consider as part of the design, where the participant copied the tabular data and pasted a Python script to clean the code. 

\subsubsection*{\textbf{Paste the data as plots}}
We observed several creative uses of \tool by participants who wanted to generate visualizations from the data. For example, participants P6 and P15 asked \tool to generate code to create a matplotlib chart. After creating a table containing summary statistics of mean and median for the sensor data, P6 asked \tool to paste the data as a Python script that can create a bar chart to show the distribution of the data with the prompt \say{can you paste the python code needed to generate a box plot using the copied data?}; \tool successfully generated the code to create the box plot. Additionally, P2, P5 asked \tool to create pie charts in Overleaf for their dataset. P5 prompted: \say{please use this data to generate a pie chart that shows the share of GDP per country}, and \tool generated a LaTeX snippet that imported the libraries and created these charts using the LaTeX \texttt{Tikz} and \texttt{pgf-pie} packages. 

\subsubsection*{\textbf{Extending the data with external knowledge}}
An unexpected application of \tool was leveraging LMM's world knowledge to extend the data during copy-paste operations. When analyzing a gene expression dataset during open exploration, P8 prompted to filter all unclassified genes that do not belong to a known gene family during paste. This classification wasn’t in the dataset, but with the LMM’s knowledge, \tool generated a regular expression to filter the data. The participant confirmed the result was accurate on a smaller sample (about 50 rows).

\subsubsection*{\textbf{Question answering}}
Another interesting application of \tool was to use it to answer questions based on the source table as opposed to pasting the original table into the new application context by P8 and P16. For example, P8 prompted \say{This is scRNAseq gene expression data, features table. How many genes are there that start with AL?}. To answer this, \tool selected the relevant column, and performed a filter operation on the tabular data, then pasted the resulting table with the one-line answer.

\paragraph{Remarks} As participants noted in their interviews, the applications they explored are often considered ``small'' tasks that yet can take considerable effort to complete. \tool's lightweight interaction approach is appealing to them: these tasks can be easy to verify, not too difficult for the LLM to complete (with the generated code), and the participants did not need to switch to other tools. Furthermore, by making easy-to-access GenAI capabilities into the copy-paste action, 
participants could easily apply \tool to new contexts that may not even have built-in AI support. 
\section{Related Work}
\paragraph{Data transfer across applications}  Based on study of information workers' routine to reuse data from various sources (databases, spreadsheets, websites, text documents, and emails), \citet{scaffidi2006games}  underscore key obstacles in data reuse comes from their needs to repair data, handle data incompatibility, and keep track of data they transferred. 
From a 90-day longitudinal study on how computer users transfer data within and across desktop applications, \citet{Woodruff2019DataTA} observed that copy-paste is the primary approach to move data comparing to other approaches like drag and drop. While commonly use, \citet{Woodruff2019DataTA} highlighted that unlike with-in app copy and pastes, users who data transfer across applications need on average 1.6 window switches to examine data being transferred, given data formatting issues. Our work incorporates data transformation and reformatting as part of the copy-paste process, to help users overcoming heterogeneous data transfer obstacles.

Data transformation plays a critical role in helping users resolve format-based compatibility issues that arise when moving data between heterogeneous applications. These transitions often introduce structural mismatches and semantic loss, requiring users to manually clean or reformat data to make it usable. For example, users may use Excel formulas to clean up data when copying from HTML webpages, or use intermediate tools or libraries like pandas~\cite{reback2020pandas}, tidyverse~\cite{wickham2019tidyverse}, Wrangler~\cite{kandel2011wrangler}, or Tableau Prep~\cite{tableauprep} to do operations that reshape, derive, and clean tabular data. However, these tools often require users to first identify the desired data shape and then perform the transformation in a separate environment, creating friction when the transformation is necessary simply to complete a task in a different application. Programming-by-example systems~\cite{gulwani2011automating} and mixed-initiative tools~\cite{kandel2011wrangler} aim to reduce user burden by inferring transformations from limited input. Data Formulator~\cite{wang2023data} builds on these foundations by embedding transformation directly within visualization authoring. Recently, with Generative AI, users who cannot or do not want to learn a new tool or programming library have resorted to using AI applications like ChatGPT, Claude, and Gemini etc. to help them perform these transformations. Yet, even with these tools, users must often bridge the gap between incompatible formats on their own. Our work reduces this burden by integrating transformation into the copy-paste process itself, automatically performing necessary conversions based on inferred context and user instructions --- eliminating the need for external tools or manual intermediate steps.n inferred context, and user instructions. 

\paragraph{Improving Clipboards}
Given the importance of clipboards for users, the first type of enhancement proposed for clipboards is to improve its contextual awareness~\cite{stolee2009revealing, stylos2004citrine} that moves beyond simple copy-paste actions but also understands the structure and intent behind them. \citet{stolee2009revealing} identify common clipboard usage patterns through empirical studies show the need for contextual awareness for several copy-paste scenarios (referred to as patterns in the paper)for smarter tool support.\citet{stylos2004citrine} propose Citrine, a system that applies intelligent transformations to clipboard data by parsing its structure and enabling context-aware pasting into various applications, such as calendar entries or spreadsheet rows. For example, users can teach Citrine how to map fields from copied data to specific application targets. Our system projects Citrine’s ideas about context and data transformation further by leveraging the capabilities of modern GenAI, which insinuate data mobility and transformations that not only adapt to previously unseen contexts but also can be guided by people’s intentions.

Among the many precursors of our work, \textit{AutoComPaste}~\cite{Zhao2012AutoComPasteAT} uses autocompletion based on previously seen documents to enhance traditional copy-paste operations. The technique offers users real-time suggestions for text that they might want to paste, based on previously copied content or frequently used text. This autocompletion is implemented by dictionary-based text matching.
In addition to enabling context-aware clipboards, Reif et al.~\cite{Reif2007SemClipO, Reif2006SemanticC} proposed the \textit{Semantic Clipboard}, designed to preserve the semantic context of the data by maintaining meaning and relationships within, thus maintaining its semantic integrity across applications. Our work shares the same motivation to use context to help maintain data semantic and structural integrity, but differentiates by its ability to accommodate user's guidance to transform the data, as well as relying on LLMs' vast knowledge representation of the world to make it robust on a variety of contexts without the need of custom plug-ins.

As prior work explores the potential of clipboards with enhanced functionalities that are context-aware, more recent work starts to explore how advances in machine learning can not only facilitate context detection but also advanced data manipulation and transformation. The \textit{Clipboard to SMILES} system~\cite{Schilter2024CMDV} shows how a smart clipboard can transform images of chemical structures into various text-based molecular representations. 
\textit{Cut-and-Paste Neural Rendering}~\cite{Bhattad2020CutandPasteNR} shows how a contextually aware clipboard can use machine learning to transform a copied image into a version that looks like it naturally belongs in a destination scene.

Recent tools have begun exploring how GenAI models can enhance universal clipboard functionality. For example, Windows PowerToys’ \texttt{Advanced Paste} allows users to paste copied data in a transformed format using a natural language prompt --- demonstrating clear user interest in such capabilities. However, \emph{Advanced Paste} remains limited to simple, prompt-based transformations within the model’s context window, which can lead to errors and hallucinations. In contrast, our system, \tool, executes code-based transformations, improving both reliability and expressiveness. Crucially, \tool also tracks contextual information about the source and destination applications --- unlike \emph{Advanced Paste}, which depends entirely on the user to specify this context. Apple’s \texttt{Writing Tools} similarly offers a GenAI-powered interface for editing text across applications, but its functionality is confined to basic rewrites such as summarization, bullet point generation, or formalization. Our work shares the same design space as these tools but pushes it further: we explore how recent advances in GenAI, such as \emph{tool use} and \emph{structured outputs}, can enable intelligent clipboards that go beyond simple rewrites --- supporting context-aware transformations, robust data handling, and the conveyance of user intent alongside content.

\section{Discussion and Future Work}

Although \tool performs above our participants expectations, its dependence on GenAI models raises the question of what other contexts, data types, and prompts will cause it to behave unexpectedly. The release cycle of different and new LLMs adds a level of design and implementation uncertainty, which perhaps calls for design principles about better error handling through lightweight human intervention. 
For example, when revealing the inner workings of \tool, many participants were interested in the option to look at the code and intermediate outputs when needed. The argument was that this could help them verify if the transformation was performed correctly, or even re-use the code in other situations. 
Future work includes exploring how to make \tool robust against unseen contexts, data, and prompts while providing ways to audit or verify its work in digestible ways.

One of \tool's current limitations is that it only works with a limited context -- the recent item in the clipboard, and a limited set of application context inferred from where the users triggered copy and paste. In the future, we can significantly improve the understanding of the implicit context of users, including the text/object context around the cursor, and other signals like user activity to better tune the context passed to the AI model. 

Our work exposes important considerations about privacy and safety. By introducing a GenAI component that can potentially move content away from a person's computer, private or unwanted information might leak. Providing a system that provides GenAI capabilities on the device is a promising way to address these situations while warning users when information might be shared outside their control.
Even in the presence of private data processing and improvements in GenAI, hallucinations and inappropriate outputs can occur, which can both disrupt the existing workflow or cause harm. Future work needs to incorporate safety measures that allow both identifying and recovering from hallucinations, as well as blocking undesired or harmful content. Multi-agent configurations are a promising approach to verify and modulate the output of GenAI.

Although our work started as an exploration of how to overcome existing challenges when people need to transfer data across different applications, it also presented how, in addition to the above, it serves as a lightweight mechanism to bring powerful GenAI capabilities to applications that do not have them. An example of this is a simple text editor, where now, by using it as both source and destination of a copy-paste operation, one effectively has an editor with GenAI capabilities like generation, summarizing, text-to-image, etc.
This type of lightweight interaction that is familiar yet can now augment the capabilities of the tools people use in affordable ways is a human-centered, compelling, and promising direction for future work.

\bibliographystyle{ACM-Reference-Format}
\bibliography{references}

@inproceedings{stylos2004citrine,
author = {Stylos, Jeffrey and Myers, Brad A. and Faulring, Andrew},
title = {Citrine: providing intelligent copy-and-paste},
year = {2004},
isbn = {1581139578},
publisher = {Association for Computing Machinery},
address = {New York, NY, USA},
url = {https://doi.org/10.1145/1029632.1029665},
doi = {10.1145/1029632.1029665},
booktitle = {Proceedings of the 17th Annual ACM Symposium on User Interface Software and Technology},
pages = {185–188},
numpages = {4},
location = {Santa Fe, NM, USA},
series = {UIST '04}
}

@article{Bhattad2020CutandPasteNR,
  title={Cut-and-Paste Neural Rendering},
  author={Anand Bhattad and David Alexander Forsyth},
  journal={ArXiv},
  year={2020},
  volume={abs/2010.05907},
  url={https://api.semanticscholar.org/CorpusID:222291281}
}

@article{Woodruff2019DataTA,
  title={Data transfer: A longitudinal analysis of clipboard and drag-and-drop use in desktop applications},
  author={Jonathan Woodruff and Jason Alexander},
  journal={Int. J. Hum. Comput. Stud.},
  year={2019},
  volume={132},
  pages={112-120},
  url={https://api.semanticscholar.org/CorpusID:201876607}
}

@article{Schilter2024CMDV,
  title={CMD + V for chemistry: Image to chemical structure conversion directly done in the clipboard},
  author={Oliver Schilter and Teodoro Laino and Philippe Schwaller},
  journal={Applied AI Letters},
  year={2024},
  url={https://api.semanticscholar.org/CorpusID:267288886}
}

@inproceedings{Reif2006SemanticC,
  title={Semantic Clipboard - Semantically Enriched Data Exchange Between Desktop Applications},
  author={Gerald Reif and Harald C. Gall and Martina Morger},
  booktitle={SemDesk},
  year={2006},
  url={https://api.semanticscholar.org/CorpusID:7907817}
}

@inproceedings{Reif2007SemClipO,
  title={SemClip - Overcoming the Semantic Gap Between Desktop Applications},
  author={Gerald Reif and Gian Marco Laube and Knud M{\"o}ller and Harald C. Gall},
  booktitle={Semantic Web Challenge},
  year={2007},
  url={https://api.semanticscholar.org/CorpusID:12402201}
}

@inproceedings{Zhao2012AutoComPasteAT,
  title={AutoComPaste: auto-completing text as an alternative to copy-paste},
  author={Shengdong Zhao and Fanny Chevalier and Wei Tsang Ooi and Chee Yuan Lee and Arpit Agarwal},
  booktitle={International Working Conference on Advanced Visual Interfaces},
  year={2012},
  url={https://api.semanticscholar.org/CorpusID:11186176}
}

@inproceedings{scaffidi2006games,
  title={Games programs play: Obstacles to data reuse},
  author={Scaffidi, Chris and Shaw, Mary and Myers, Brad},
  booktitle={2nd Workshop on End User Software Engineering},
  year={2006}
}

@inproceedings{stolee2009revealing,
  title={Revealing the copy and paste habits of end users},
  author={Stolee, Kathryn T and Elbaum, Sebastian and Rothermel, Gregg},
  booktitle={2009 IEEE Symposium on Visual Languages and Human-Centric Computing (VL/HCC)},
  pages={59--66},
  year={2009},
  organization={IEEE}
}

@inproceedings{graham2008probes,
  title={Probes and participation},
  author={Graham, Connor and Rouncefield, Mark},
  booktitle={Proceedings of the Tenth Anniversary Conference on Participatory Design 2008},
  pages={194--197},
  year={2008}
}

@inproceedings{boehner2007hci,
  title={How HCI interprets the probes},
  author={Boehner, Kirsten and Vertesi, Janet and Sengers, Phoebe and Dourish, Paul},
  booktitle={Proceedings of the SIGCHI conference on Human factors in computing systems},
  pages={1077--1086},
  year={2007}
}

@inproceedings{hutchinson2003technology,
  title={Technology probes: inspiring design for and with families},
  author={Hutchinson, Hilary and Mackay, Wendy and Westerlund, Bo and Bederson, Benjamin B and Druin, Allison and Plaisant, Catherine and Beaudouin-Lafon, Michel and Conversy, St{\'e}phane and Evans, Helen and Hansen, Heiko and others},
  booktitle={Proceedings of the SIGCHI conference on Human factors in computing systems},
  pages={17--24},
  year={2003}
}

@article{gaver1999design,
  title={Design: cultural probes},
  author={Gaver, Bill and Dunne, Tony and Pacenti, Elena},
  journal={interactions},
  volume={6},
  number={1},
  pages={21--29},
  year={1999},
  publisher={ACM New York, NY, USA}
}

@article{zhang2020data,
  title={How do data science workers collaborate? roles, workflows, and tools},
  author={Zhang, Amy X and Muller, Michael and Wang, Dakuo},
  journal={Proceedings of the ACM on Human-Computer Interaction},
  volume={4},
  number={CSCW1},
  pages={1--23},
  year={2020},
  publisher={ACM New York, NY, USA}
}

@MISC{2024genericballot,
  title        = "Generic ballot Polls",
  author       = "Best, Ryan and Bycoffe, Aaron and King, Ritchie and Mehta,
                  Dhrumil and Wiederkehr, Anna",
  booktitle    = "FiveThirtyEight",
  month        =  jun,
  year         =  2018,
  howpublished = "\url{https://web.archive.org/web/20240823073303/https://projects.fivethirtyeight.com/polls/generic-ballot/}",
  note         = "Accessed: 2024-9-10",
  language     = "en"
}

@MISC{2024presential,
  title        = "National : President: general election : 2024 Polls",
  author       = "Best, Ryan and Bycoffe, Aaron and King, Ritchie and Mehta,
                  Dhrumil and Wiederkehr, Anna",
  booktitle    = "FiveThirtyEight",
  month        =  jun,
  year         =  2018,
  howpublished = "\url{https://web.archive.org/web/20240823040520/https://projects.fivethirtyeight.com/polls/president-general/2024/national/}",
  note         = "Accessed: 2024-9-10",
  language     = "en"
}

@MISC{2020olympics,
  title        = "Tokyo 2020 Olympic Athletes - Biographies, Medals \& More",
  author       = "{IOC}",
  booktitle    = "Olympics.com",
  howpublished = "\url{https://web.archive.org/web/20240729152542/https://olympics.com/en/olympic-games/tokyo-2020/athletes}",
  note         = "Accessed: 2024-9-10",
  language     = "en"
}

@inproceedings{beaudouin2000instrumental,
  title={Instrumental interaction: an interaction model for designing post-WIMP user interfaces},
  author={Beaudouin-Lafon, Michel},
  booktitle={Proceedings of the SIGCHI conference on Human factors in computing systems},
  pages={446--453},
  year={2000}
}

@book{holtzblatt1997contextual,
  title={Contextual design: defining customer-centered systems},
  author={Holtzblatt, Karen and Beyer, Hugh},
  year={1997},
  publisher={Elsevier}
}

@article{beaudouin2021generative,
  title={Generative theories of interaction},
  author={Beaudouin-Lafon, Michel and B{\o}dker, Susanne and Mackay, Wendy E},
  journal={ACM Transactions on Computer-Human Interaction (TOCHI)},
  volume={28},
  number={6},
  pages={1--54},
  year={2021},
  publisher={ACM New York, NY}
}

@MISC{openai-assistant,
  title        = "{OpenAI} Platform",
  howpublished = "\url{https://platform.openai.com/docs/assistants/overview}",
  note         = "Accessed: 2024-9-12",
  language     = "en"
}

@MISC{gpt4o,
  title        = "{GPT}-{4o}",
  author       = "{Wikipedia contributors}",
  booktitle    = "Wikipedia, The Free Encyclopedia",
  publisher    = "Wikimedia Foundation, Inc.",
  month        =  sep,
  year         =  2024,
  howpublished = "\url{https://en.wikipedia.org/wiki/GPT-4o}"
}

@MISC{gpt4o-mini,
  title        = "{GPT}-{4o} mini: advancing cost-efficient intelligence",
  howpublished = "\url{https://openai.com/index/gpt-4o-mini-advancing-cost-efficient-intelligence/}",
  note         = "Accessed: 2024-9-12",
  language     = "en"
}

@MISC{electron-clipboard,
  title        = "clipboard",
  howpublished = "\url{https://electronjs.org/docs/latest/api/clipboard}",
  note         = "Accessed: 2024-9-12",
  language     = "en"
}

@MISC{office-copilot,
  title        = "Introducing Microsoft 365 Copilot – your copilot for work",
  author       = "Spataro, Jared",
  booktitle    = "The Official Microsoft Blog",
  month        =  mar,
  year         =  2023,
  howpublished = "\url{https://blogs.microsoft.com/blog/2023/03/16/introducing-microsoft-365-copilot-your-copilot-for-work/}",
  note         = "Accessed: 2024-9-12",
  language     = "en"
}

@article{reback2020pandas,
  title={pandas-dev/pandas: Pandas 1.0. 5},
  author={Reback, Jeff and McKinney, Wes and Van Den Bossche, Joris and Augspurger, Tom and Cloud, Phillip and Klein, Adam and Hawkins, Simon and Roeschke, Matthew and Tratner, Jeff and She, Chang and others},
  journal={Zenodo},
  year={2020}
}

@article{wickham2019tidyverse,
  title={Welcome to the Tidyverse},
  author={Wickham, Hadley and Averick, Mara and Bryan, Jennifer and Chang, Winston and McGowan, Lucy D'Agostino and Fran{\c{c}}ois, Romain and Grolemund, Garrett and Hayes, Alex and Henry, Lionel and Hester, Jim and others},
  journal={Journal of open source software},
  volume={4},
  number={43},
  pages={1686},
  year={2019}
}

@inproceedings{kandel2011wrangler,
  title={Wrangler: Interactive visual specification of data transformation scripts},
  author={Kandel, Sean and Paepcke, Andreas and Hellerstein, Joseph and Heer, Jeffrey},
  booktitle={Proceedings of the sigchi conference on human factors in computing systems},
  pages={3363--3372},
  year={2011}
}

@MISC{tableauprep,
  title        = "Tableau Prep",
  booktitle    = "Tableau",
  howpublished = "\url{https://www.tableau.com/products/prep}",
  note         = "Accessed: 2025-4-9",
  language     = "en"
}

@article{gulwani2011automating,
  title={Automating string processing in spreadsheets using input-output examples},
  author={Gulwani, Sumit},
  journal={ACM Sigplan Notices},
  volume={46},
  number={1},
  pages={317--330},
  year={2011},
  publisher={ACM New York, NY, USA}
}

@article{wang2023data,
  title={Data formulator: Ai-powered concept-driven visualization authoring},
  author={Wang, Chenglong and Thompson, John and Lee, Bongshin},
  journal={IEEE Transactions on Visualization and Computer Graphics},
  volume={30},
  number={1},
  pages={1128--1138},
  year={2023},
  publisher={IEEE}
}

\appendix
\pagebreak

\section{System usability scale questionnaire and NASA TLX measures}

\begin{table}[h]\small
\begin{tabularx}{\columnwidth}{>{\hsize=1\hsize}X} \hline
Q1.1 It was easy to complete the task using the tool provided (1 - Strongly Disagree, 7 - Strongly Agree) \\
Q1.2 I am happy with the final system output (1 - Strongly Disagree, 7 - Strongly Agree) \\
Q1.3 Other people will rate the final output of the system highly (1 - Strongly Disagree, 7 - Strongly Agree) \\
Q1.4 I could easily communicate my creative goals (1 - Strongly Disagree, 7 - Strongly Agree) \\
Q1.5 I could easily plan and manage the many creative tasks. (1 - Strongly Disagree, 7 - Strongly Agree) \\
Q1.6 I could easily maintain control of my creative process. (1 - Strongly Disagree, 7 - Strongly Agree) \\
Q1.7 I could explore the creative space using the system. (1 - Strongly Disagree, 7 - Strongly Agree) \\
Q1.8 I could easily iterate, revise, and refine using the system. (1 - Strongly Disagree, 7 - Strongly Agree) \\
Q1.9 I am more confident in my creative skills after having used the system. (1 - Strongly Disagree, 7 - Strongly Agree) \\
Q1.10 I have a strong sense of ownership of the creative outcome. (1 - Strongly Disagree, 7 - Strongly Agree) \\
Q1.11 The system allowed me to reflect on my creative process. (1 - Strongly Disagree, 7 - Strongly Agree) \\
Q1.12 I trust the information generated by the system. (1 - Strongly Disagree, 7 - Strongly Agree) \\
\\
Q2.1. How mentally demanding was this task with this tool? (1—Very Low, 7—Very High)\\
Q2.2. How hurried or rushed were you during this task? (1—Very Low, 7—Very High)\\
Q2.3. How successful would you rate yourself in accomplishing this task? (1—Perfect, 7—Failure)\\
Q2.4. How hard did you have to work to accomplish your level of performance? (1—Very Low, 7—Very High)\\
Q2.5. How insecure, discouraged, irritated, stressed, and annoyed were you? (1—Very Low, 7—Very High)\\\hline
\end{tabularx}
\caption{After design probe tasks, participants answered the system usability scale (questions 1.1 - 1.12; 7-point Likert Scale) and the subjective workload using NASA TLX measures (questions 2.1 - 2.5; 7-point Likert Scale).}
\Description{Shows a table containing the questions for the system usability scale and the NASA TLX workload assessment.}
\label{tab:posttask}
\end{table}

\begin{figure*}[t]
    \centering
    \includegraphics[width=0.58\linewidth]{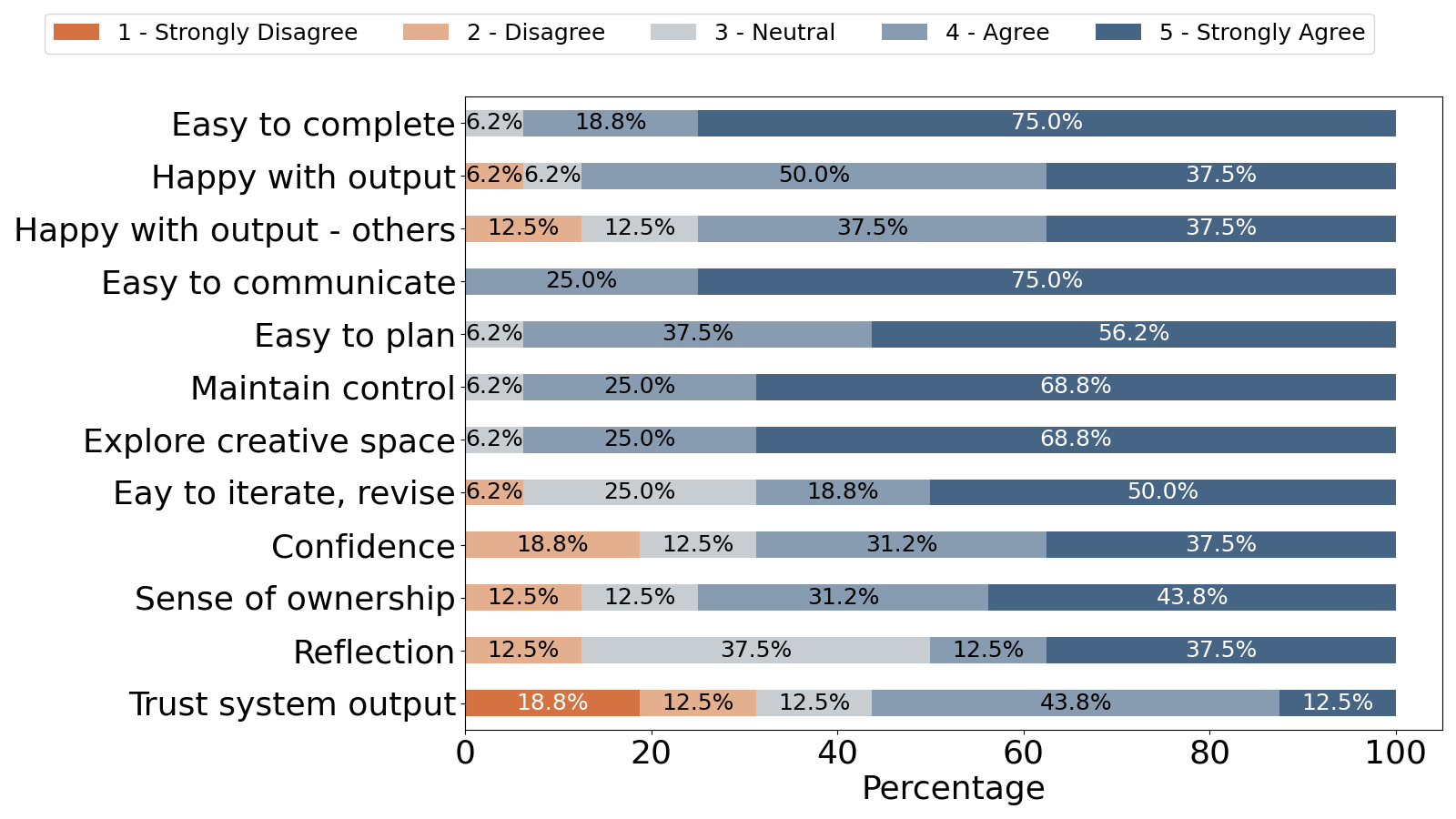}
    \includegraphics[width=0.35\linewidth]{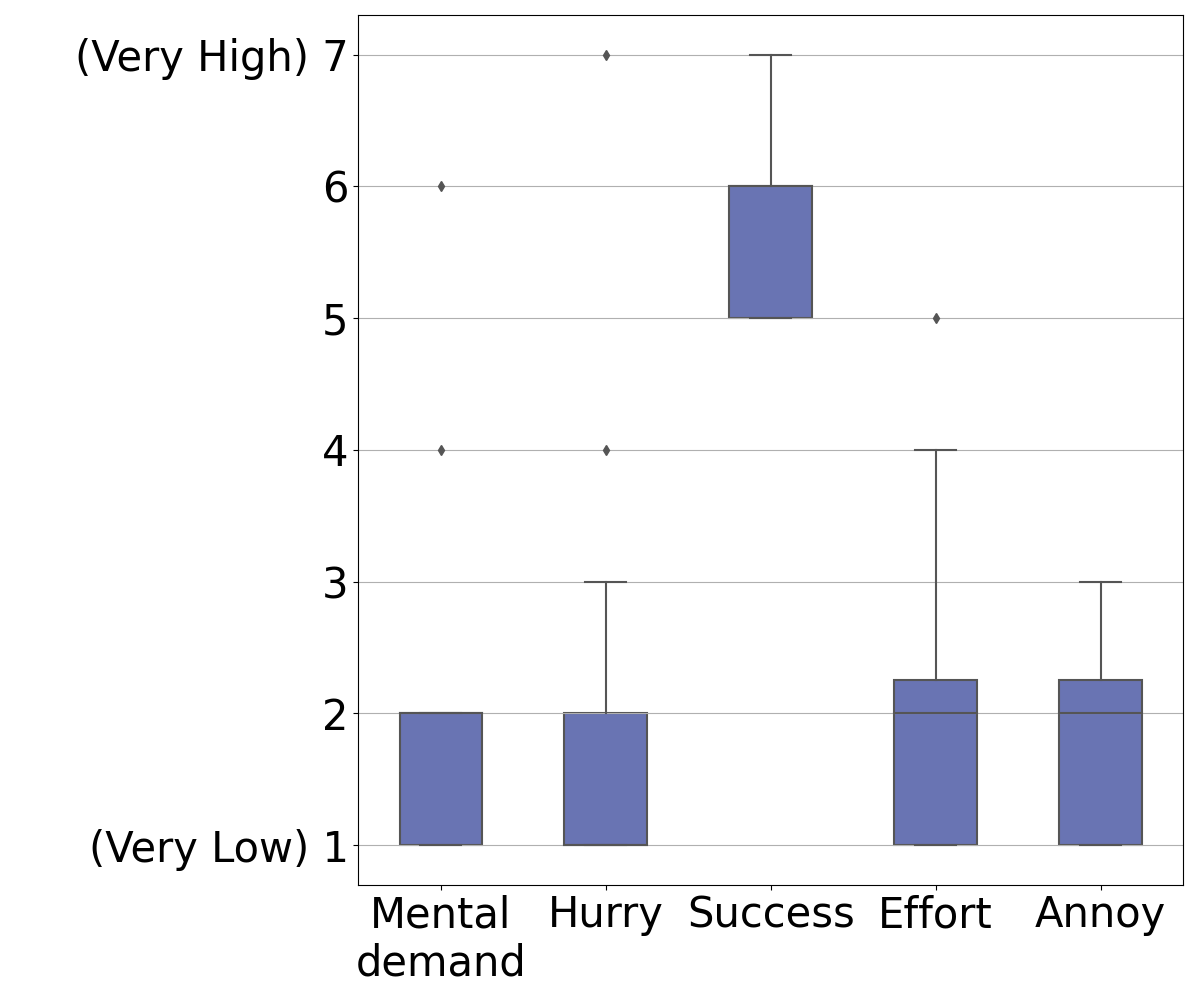}
    \caption{Participant responses for the System usability scale, and NATA TLX questionnaire.}
    \Description{Shows a chart plotting the participant responses for the System usability scale, and NATA TLX questionnaire.}
    \label{fig:tlx-and-experience}
\end{figure*}

\section{Tools provided to \tool Agent}~\label{app:agenttools}

\begin{itemize}
    \item \textit{GetClipboardSummary}: When invoked by the agent, this tool returns the summary of the clipboard data, which includes the type, source of the data, and a sample of the raw data (truncated to 10000 characters).
    \item \textit{AddStructuredDataUsingCode}: This tool takes a Python function generated by the LMM model, executes it in the sandbox environment, and adds the result to \splitt{ContextObject['structuredData']}. The entire \texttt{ContextObject} is passed to the function as a parameter.
    \item \textit{AddMetaDataUsingCode}: Similar to above, this tool takes a Python function, executes it and stores the result in \splitt{ContextObject['metadata']['key']}. 'key' is provided by the model.  
    \item \texttt{sampleContextObject}
    This tool is for the agent to generate and execute Python code to sample, poll, or access parts of the data to understand the data context. The entire clipboard object is passed as an input parameter to the function.
    \item \textit{AddTransformationUsingCode}: Similar to above, this is used to generate a transformation by executing code and storing it in \splitt{ContextObject['transformations']['key']}. 'key' is provided by the model.
    \item \texttt{runPythonCode}: This tool executes a Python function in the sandbox environment and returns the output object. The entire clipboard context object is passed as a parameter to this function.
    \item \textit{writeToTempFile}: The agent can use this to write one of the transformations to a temporary file. The tool accepts a \texttt{key}, and a file extension \texttt{ext}. The tool will store the contents of \splitt{ContextObject['transformations']['key']} to \texttt{<tempfile>.ext}.
    \item \texttt{pasteToDestination}: This tool takes two parameters: \texttt{key} and \texttt{contentType}. The content type can either be text, HTML, or RTF. This tool will paste the contents of \splitt{ContextObject['transformations']['key']} to the destination app. In case the paste to the destination app fails, the tool will insert the contents to the OS clipboard for the user to trigger the paste manually. 
\end{itemize}

\end{document}